\documentstyle{article}

\newcommand{\ba}{\begin{eqnarray}}
\newcommand{\ea}{\end{eqnarray}}
\begin{document}
\pagestyle{plain}

\title{Algebraic models of hadron structure:\\
I. Nonstrange baryons}
\author{R.~Bijker\\
R.J.~Van de Graaff Laboratory, University of Utrecht,\\
P.O. Box 80000, 3508 TA Utrecht, The Netherlands
\and
F.~Iachello\\
Center for Theoretical Physics, Sloane Laboratory,\\
Yale University, New Haven, CT 06511, U.S.A.
\and
A.~Leviatan\\
Racah Institute of Physics, The Hebrew University,\\
Jerusalem 91904, Israel}
\date{}
\maketitle

\begin{abstract}
We introduce an algebraic framework for the description of baryons.
Within this framework we study a collective string-like model
and show that this model gives a good overall description of
the presently available data.
We discuss in particular masses and electromagnetic couplings,
including the transition form factors that can be measured at new
electron facilities.
\end{abstract}

\section{Introduction}

In the last 20 years QCD has emerged as the theory of strong interactions.
This theory has been tested in the perturbative regime by several
experiments at CERN and other laboratories. However, in the nonperturbative
regime, defined in this article as
$E \displaystyle\mathop{<}_{\sim} 3$
GeV, no solution of QCD is
known, except for lattice calculations of the ground state and its
properties. Even with the development of new dedicated computers, the
lattice calculation of the excitation spectrum of hadrons is a daunting
problem and its solution is still years away.

In a series of papers starting with this one we address the problem of
the spectroscopy of baryons in the nonperturbative regime using methods
introduced in this field 30 years ago to describe its flavor-spin part
\cite{sf} and extended here to include the space part. This extension is
stimulated by the success that the use of algebraic methods has had in
other fields of physics, most notably nuclear \cite{ibm} and
molecular \cite{vibron} physics. By making use of these methods we are
able to calculate in a straigthforward way all observable quantities
and thus test various models of hadronic structure.
In particular, we are able to contrast the nonrelativistic \cite{IK} or
relativized \cite{CI} valence quark models with string-like models.
In doing so, we also provide a more transparent understanding
of the extent to which the nonrelativistic reduction of the
perturbative one-gluon exchange interaction used in \cite{IK,CI}
holds in the nonperturbative regime.
However, the main purpose of this article is not so much that of
testing well-known models, but rather that of (i) introducing the
algebraic framework and, most importantly, of (ii) studying a new
`collective' model of baryons which appears to be give a realistic
description {\it vis \`a vis} the experimental data,
especially in the form factors.

The outline of this article is as follows: in Sections~2--4 we
present some general properties of the algebraic approach, in
Sections~5 and~6 we discuss the mass spectrum and in Sections~7--10
the electromagnetic couplings of nonstrange baryons.
Some of the more technical details are presented in the appendices.

\section{Models of baryons}
\setcounter{equation}{0}

We consider baryons to be built of three constituent parts. The global
internal quantum numbers of these three parts are taken to be:
flavor=triplet=u,d,s; spin=doublet=1/2; and color=triplet (we
do not consider here heavy quark flavors). The internal algebraic
structure of the constituent parts is thus the usual
\ba
{\cal G}_i &=& SU_{f}(3) \otimes SU_{s}(2)
\otimes SU_{c}(3) ~.
\ea
(We use in this paper the notation appropriate to groups,
{\it i.e.} upper case letters and $\otimes$ signs, rather than to
algebras, {\it i.e.} lower case letters and $\oplus$ signs.)
The quantum numbers need not be all concentrated at one point
but may be distributed over a volume of size equal to the hadronic
size. In particular, in this paper we discuss mostly the string-like
configuration depicted in Figure~\ref{geometry}. Although the string
is fat, we shall, from time to time, idealize it as a thin string
with a distribution of mass, charge and magnetic moments.
The algebraic method that we shall introduce can easily be applied to
the single-particle valence quark model as well and therefore we shall
present also results for this model when needed for comparison.

The main purpose of this paper is to study the properties of baryon
resonances in terms of a string-like model (see Figure~\ref{geometry}),
in particular the mass spectrum and electromagnetic couplings.
In order to do so, we need a framework within which
this study can be done. In the valence quark model this is usually a
Schr\"{o}dinger-like differential equation with two-body interactions.
We prefer here instead to use a more general method based on a
bosonic quantization of the relevant degrees of freedom. This method
has been extensively used in nuclear and molecular physics \cite{books}.
The relevant degrees of freedom characterizing the configuration in
Figure~\ref{geometry} are the two Jacobi coordinates $\vec{\rho}$,
$\vec{\lambda}$ (in addition to the center-of-mass coordinate
which is not relevant for the excitation spectrum).
The relative Jacobi coordinates are given by
\ba
\vec{\rho} &=& \frac{1}{\sqrt{2}}
(\vec{r}_1 - \vec{r}_2) ~,
\nonumber\\
\vec{\lambda} &=& \frac{1}{\sqrt{6}}
(\vec{r}_1 + \vec{r}_2 - 2\vec{r}_3) ~, \label{jacobi}
\ea
where $\vec{r}_1$, $\vec{r}_2$ and $\vec{r}_3$ denote the end points of
the string configuration in Figure~\ref{geometry}.
The method of bosonic quantization consists in introducing two vector
boson operators (one for each relative coordinate) which are
related to the coordinates, $\vec{\rho}$ and $\vec{\lambda}$, and their
conjugate momenta, $\vec{p}_{\rho}$ and $\vec{p}_{\lambda}$, by
\ba
b^{\dagger}_{\rho,m} &=&
\frac{1}{\sqrt{2}} ( \rho_m - i \, p_{\rho,m} ) ~,
\nonumber\\
b_{\rho,m} &=&
\frac{1}{\sqrt{2}} ( \rho_m + i \, p_{\rho,m} ) ~,
\nonumber\\
b^{\dagger}_{\lambda,m} &=&
\frac{1}{\sqrt{2}} ( \lambda_m - i \, p_{\lambda,m} ) ~,
\nonumber\\
b_{\lambda,m} &=&
\frac{1}{\sqrt{2}} ( \lambda_m + i \, p_{\lambda,m} ) ~, \label{boson}
\ea
with $m=-1,0,1$, and an additional auxiliary scalar boson, $s^{\dagger}$,
$s$. These operators satisfy usual boson commutation relations and
operators of different type commute. If we denote generically the set
of seven creation operators by $c^{\dagger}_{\alpha}$,
\ba
b^{\dagger}_{\rho,m} ~, \; b^{\dagger}_{\lambda,m} ~, \;
s^{\dagger} &\equiv& c^{\dagger}_{\alpha} ~,
\ea
with $\alpha=1,\ldots,7$, the bilinear products
\ba
{\cal G}_r \;:\; G_{\alpha \alpha^{\prime}}
\;=\; c^{\dagger}_{\alpha} c_{\alpha^{\prime}} ~,
\ea
with $\alpha,\alpha^{\prime}=1,\ldots,7$ generate the Lie algebra of
$U(7)$. This bosonic quantization scheme follows the usual prescription
that any problem in $\nu$ space degrees of freedom be written in terms of
elements of the Lie algebra $U(\nu+1)$, and that all its states be assigned
to the totally symmetric representation $[N]$ of $U(\nu+1)$ \cite{FI}.
All operators can be expanded into elements of ${\cal G}_r = U(7)$, and
states can be constructed by acting with the boson operators on a vacuum
\ba
\frac{1}{{\cal N}} \, (b^{\dagger}_{\rho})^{n_{\rho}}
(b^{\dagger}_{\lambda})^{n_{\lambda}}
(s^{\dagger})^{N-n_{\rho}-n_{\lambda}} \, |0 \rangle ~,
\ea
where ${\cal N}$ is a normalization factor.
The complete algebraic structure of the problem is thus
\ba
{\cal G} \;=\; {\cal G}_r \otimes {\cal G}_i \;=\;
U(7) \otimes SU_{f}(3) \otimes SU_{s}(2)
\otimes SU_{c}(3) ~.
\ea
${\cal G}$ is the spectrum generating algebra (SGA) of baryon
structure.

\section{The algebra of U(7)}
\setcounter{equation}{0}

We want to construct states and operators that transform according
to irreducible representations of the rotation group (since the problem
is rotationally invariant). The creation operators, $b^{\dagger}_{\rho}$
and $b^{\dagger}_{\lambda}$, transform by definition as
vectors under rotation. The annihilation operators do not. It is easy
to construct operators that transform appropriately. They are
\ba
\tilde{b}_{\rho,m} &=& (-1)^{1-m} b_{\rho,-m} ~,
\nonumber\\
\tilde{b}_{\lambda,m} &=& (-1)^{1-m} b_{\lambda,-m} ~,
\nonumber\\
\tilde{s} &=& s ~.
\ea
Using these operators one can rewrite the 49 elements of $U(7)$ in
their angular-momentum coupled Racah form
\ba
( c^{\dagger}_l \times \tilde{c}_{l^{\prime}} )^{(L)}_M
&=& \sum_{m,m^{\prime}} \langle l,m,l^{\prime},m^{\prime} | L,M \rangle
\, c^{\dagger}_{l,m} \tilde{c}_{l^{\prime},m^{\prime}} ~,
\ea
where $c_l^{\dagger}$ ($\tilde{c}_l$) for $l=1$ denotes the vector bosons
$b^{\dagger}_{\rho}$ ($\tilde{b}_{\rho}$), $b^{\dagger}_{\lambda}$
($\tilde{b}_{\lambda}$) and for $l=0$ the scalar boson $s^{\dagger}$
($\tilde{s}$). The cross denotes a tensor product with
respect to $SO(3)$.

Another unavoidable problem in baryon structure is that, if some of
the constituent parts are identical, one must construct states and
operators that transform according to representations of the
permutation group (either $S_3$ for three identical parts or $S_2$
for two identical parts). The problem is particularly acute for
nonstrange baryons, if one assumes that the three constituent parts
are identical. In this case one must construct operators that transform
as irreducible representations of $S_3$. In this construction \cite{BL},
we use the transposition $P(12)$ and the cyclic permutation $P(123)$.
All other permutations can be expressed in terms of these two
elementary ones. The transformation properties under $S_3$ of all
operators of interest follow from those of the building blocks.
Using the definitions of Eqs.~(\ref{jacobi}) and~(\ref{boson}), one has
the following transformation properties of the creation operators
under $S_3$
\ba
P(12) \left( \begin{array} {l} s^{\dagger} \\ b^{\dagger}_{\rho,m} \\
b^{\dagger}_{\lambda,m} \end{array} \right) &=&
\left( \begin{array}{rrr} 1 & 0 & 0 \\ 0 & -1 & 0 \\ 0 & 0 & 1
\end{array} \right) \left( \begin{array} {l} s^{\dagger} \\
b^{\dagger}_{\rho,m} \\ b^{\dagger}_{\lambda,m} \end{array} \right) ~,
\nonumber\\
\nonumber\\
P(123) \left( \begin{array} {l} s^{\dagger} \\ b^{\dagger}_{\rho,m} \\
b^{\dagger}_{\lambda,m} \end{array} \right) &=&
\left( \begin{array}{ccc} 1 & 0 & 0 \\
0 &  \cos (2\pi/3) &  \sin (2\pi/3) \\
0 & -\sin (2\pi/3) &  \cos (2\pi/3) \end{array} \right)
\left( \begin{array} {l} s^{\dagger} \\ b^{\dagger}_{\rho,m} \\
b^{\dagger}_{\lambda,m} \end{array} \right) ~. \label{s3}
\ea
There are three different symmetry classes for the permutation of
three objects:
\ba
\begin{array}{lcl}
\begin{array} {l} \Box \Box \Box \end{array} &\equiv&
S(\mbox{ymmetric}) \\
& & \\
\begin{array} {l} \Box \Box \\ \Box \end{array} &\equiv&
M(\mbox{ixed symmetry}) \\
& & \\
\begin{array} {l} \Box \\ \Box \\ \Box \end{array} &\equiv&
A(\mbox{ntisymmetric}) \end{array}
\ea
with dimensions 1, 2 and 1, respectively. Eq.~(\ref{s3}) shows that the
scalar boson, $s^{\dagger}$, transforms as the symmetric representation,
$S$, while the two vector bosons, $b^{\dagger}_{\rho}$ and
$b^{\dagger}_{\lambda}$, transform as the two components, $M_{\rho}$
and $M_{\lambda}$, of the mixed symmetry representation.
Alternatively, the three symmetry classes can
be labeled by the irreducible representations of the point group
$D_{3}$ (which is isomorphic to $S_3$) as $A_{1}$, $E$ and $A_2$,
respectively.

Now we can rewrite the 49 elements of the algebra of $U(7)$ in terms
of the operators that transform as irreducible representations
of $SO(3)$ and $S_3$,
\ba
\hat D_{\rho,m} &=& (b^{\dagger}_{\rho} \times \tilde{s} -
s^{\dagger} \times \tilde{b}_{\rho})^{(1)}_m ~,
\nonumber\\
\hat D_{\lambda,m} &=& (b^{\dagger}_{\lambda} \times \tilde{s} -
s^{\dagger} \times \tilde{b}_{\lambda})^{(1)}_m ~,
\nonumber\\
\hat A_{\rho,m} &=& i \, (b^{\dagger}_{\rho} \times \tilde{s} +
s^{\dagger} \times \tilde{b}_{\rho})^{(1)}_m ~,
\nonumber\\
\hat A_{\lambda,m} &=& i \, (b^{\dagger}_{\lambda} \times \tilde{s} +
s^{\dagger} \times \tilde{b}_{\lambda})^{(1)}_m ~,
\nonumber\\
\hat G^{(l)}_{M_{\rho},m} &=&
( b^{\dagger}_{\rho} \times \tilde{b}_{\lambda}
+ b^{\dagger}_{\lambda} \times \tilde{b}_{\rho} )^{(l)}_m ~,
\nonumber\\
\hat G^{(l)}_{M_{\lambda},m} &=&
( b^{\dagger}_{\rho} \times \tilde{b}_{\rho}
- b^{\dagger}_{\lambda} \times \tilde{b}_{\lambda} )^{(l)}_m ~,
\nonumber\\
\hat G^{(l)}_{S,m} &=& ( b^{\dagger}_{\rho} \times \tilde{b}_{\rho}
+ b^{\dagger}_{\lambda} \times \tilde{b}_{\lambda} )^{(l)}_m ~,
\nonumber\\
\hat G^{(l)}_{A,m} &=& i \,
( b^{\dagger}_{\rho} \times \tilde{b}_{\lambda}
- b^{\dagger}_{\lambda} \times \tilde{b}_{\rho} )^{(l)}_m ~,
\nonumber\\
\hat{n}_s &=& (s^{\dagger} \times \tilde{s})^{(0)}_0 ~, \label{gen}
\ea
with $l=0,1,2$. For future reference, we present here the explicit
expressions of some other operators of interest. They are all linear
combinations of the generators of Eq.~(\ref{gen})
\ba
\hat n_{\rho} &=& \sqrt{3} \,
( b^{\dagger}_{\rho} \times \tilde{b}_{\rho} )^{(0)}_0 ~,
\nonumber\\
\hat n_{\lambda} &=& \sqrt{3} \,
( b^{\dagger}_{\lambda} \times \tilde{b}_{\lambda} )^{(0)}_0 ~,
\nonumber\\
\hat N &=& \hat n_s + \hat n_{\rho} + \hat n_{\lambda} ~,
\nonumber\\
\hat L_m &=& \sqrt{2} \, \hat G^{(1)}_{S,m} ~,
\nonumber\\
\hat K_y &=& - \sqrt{3} \, \hat G^{(0)}_{A,0} ~. \label{operators}
\ea
Also for future reference and in view of the fact that this
problem is of relevance to other fields as well (triatomic molecules
\cite{BDL}), we summarize in Table~\ref{sthree} the transformation
properties of some linear and bilinear combinations of boson operators.

\section{Basis states}
\setcounter{equation}{0}

In order to do calculations one needs to construct a complete set of
basis states for the representations of $U(7)$. These are obtained
by considering subalgebras of $U(7)$.
There are two bases of particular interest:
\ba
U(7) \supset U_{\rho}(3) \otimes U_{\lambda}(4) \supset
\left\{ \begin{array}{c} U_{\rho}(3) \otimes U_{\lambda}(3) \\ \\
U_{\rho}(3) \otimes SO_{\lambda}(4) \end{array} \right\}
\supset SO_{\rho}(3) \otimes SO_{\lambda}(3) \supset SO(3)
\supset SO(2) ~.
\ea

The first one corresponds to two coupled three-dimensional harmonic
oscillators. The states in this basis are characterized by a set of
quantum numbers related to the irreducible
representations of the subgroups
\ba
\left| \begin{array}{ccccccccc}
U(7) & \supset & U_{\rho}(3) & \otimes & U_{\lambda}(4)
& \supset & U_{\rho}(3) & \otimes & U_{\lambda}(3) \\
\downarrow & & \downarrow & & & & & & \downarrow \\
N & & n_{\rho} & & & & & & n_{\lambda} \end{array} \right. \hspace{2cm}
\nonumber\\
\left. \begin{array}{cccccccc}
\supset & SO_{\rho}(3) & \otimes & SO_{\lambda}(3) & \supset & SO(3)
& \supset & SO(2) \\
& \downarrow & & \downarrow & & \downarrow & & \downarrow \\
& L_{\rho} & & L_{\lambda} & & L & & M_L \end{array} \right> ~.
\ea
For a given value of $N$, {\it i.e} the model space in which calculations
are done, one has
\ba
n_{\rho} &=& 0,1,\ldots,N ~,
\nonumber\\
n_{\lambda} &=& 0,1,\ldots,N-n_{\rho} ~,
\nonumber\\
L_{\rho} &=& n_{\rho},n_{\rho}-2,\ldots,1 \mbox{ or } 0 ~,
\nonumber\\
L_{\lambda} &=& n_{\lambda},n_{\lambda}-2,\ldots,1 \mbox{ or } 0 ~,
\nonumber\\
L &=& |L_{\rho}-L_{\lambda}|,|L_{\rho}-L_{\lambda}|+1,\ldots,
L_{\rho}+L_{\lambda} ~,
\nonumber\\
M_L &=& -L,-L+1,\ldots,L ~.
\ea
The parity of the state is $\pi=(-)^{L_{\rho}+L_{\lambda}}$.
The basis states are then uniquely labeled by
\ba
|N,(n_{\rho},L_{\rho}),(n_{\lambda},L_{\lambda});L,M_L \rangle ~.
\label{basis1}
\ea
The same basis of two coupled harmonic oscillators is employed in the
nonrelativistic and relativized quark models. Early quark model
calculations \cite{IK} used $n_{\rho}+n_{\lambda} \leq 2$,
while more recent calculations \cite{CI} have used
$n_{\rho}+n_{\lambda} \leq 6$. These choices correspond in $U(7)$
to taking the total number of bosons equal to $N=2$ and $N=6$,
respectively.

The second basis is very convenient to evaluate matrix elements
of the electromagnetic transition operator (see Appendix~D).
It corresponds to a coupled system of a three-dimensional
harmonic oscillator, $U_{\rho}(3)$, and a three-dimensional Morse
oscillator, $SO_{\lambda}(4)$. The states in this basis are labeled by
\ba
\left| \begin{array}{ccccccccc}
U(7) & \supset & U_{\rho}(3) & \otimes & U_{\lambda}(4)
& \supset & U_{\rho}(3) & \otimes & SO_{\lambda}(4) \\
\downarrow & & \downarrow & & & & & & \downarrow \\
N & & n_{\rho} & & & & & & \omega \end{array} \right. \hspace{2cm}
\nonumber\\
\left. \begin{array}{cccccccc}
\supset & SO_{\rho}(3) & \otimes & SO_{\lambda}(3) & \supset & SO(3)
& \supset & SO(2) \\
& \downarrow & & \downarrow & & \downarrow & & \downarrow \\
& L_{\rho} & & L_{\lambda} & & L & & M_L \end{array} \right> ~.
\ea
For a given value of $N$ the allowed values of the quantum numbers
are given by
\ba
\omega &=& N-n_{\rho},N-n_{\rho}-2,\ldots,1 \mbox{ or } 0 ~,
\nonumber\\
L_{\lambda} &=& 0,1,\ldots,\omega ~.
\ea
The allowed values of $n_{\rho}$, $L_{\rho}$, $L$ and $M_L$ are the
same as in the harmonic oscillator basis.
The parity is $\pi=(-)^{L_{\rho}+L_{\lambda}}$.
Summarizing, the basis states are uniquely labeled by
\ba
|N,(n_{\rho},L_{\rho}),(\omega,L_{\lambda});L,M_L \rangle ~.
\label{basis2}
\ea

\section{Mass operator}
\setcounter{equation}{0}

In general the mass operator depends both on the spatial and the
internal degrees of freedom. We first discuss the contribution from
the spatial part and then that of the spin-flavor part.

\subsection{Space part}

The algebraic framework of the previous sections allows one to study
the excitation spectra of objects with the geometric shape of
Figure~\ref{geometry}.
In nonrelativistic problems the spectrum is obtained by expanding the
hamiltonian in terms of operators of the algebra ${\cal G}_r$. Here we
prefer to expand the mass-squared operator into elements of ${\cal G}_r$,
\ba
\hat M^2 &=& f(G_{\alpha \alpha^{\prime}})~, \hspace{1cm}
G_{\alpha \alpha^{\prime}} \; \in \; {\cal G}_r ~.
\ea
The expansion is usually a polynomial in $G_{\alpha \alpha^{\prime}}$.
When the three constituent parts are identical, the mass-squared operator
must transform as the symmetric representation $S$ (or $A_1$) of $S_3$
(or $D_3$). The most general $\hat M^2$ operator that preserves angular
momentum and parity, transforms as a scalar under the permutation group,
and is at most quadratic in $G_{\alpha \alpha^{\prime}}$ is
\ba
\hat M^2 &=& M^2_0 + \epsilon_s \, s^{\dagger} \tilde{s}
- \epsilon_p \, (b_{\rho}^{\dagger} \cdot \tilde{b}_{\rho}
+ b_{\lambda}^{\dagger} \cdot \tilde{b}_{\lambda})
+ u_0 \, (s^{\dagger} s^{\dagger} \tilde{s} \tilde{s}) - u_1 \,
  s^{\dagger} ( b^{\dagger}_{\rho} \cdot \tilde{b}_{\rho}
+ b^{\dagger}_{\lambda} \cdot \tilde{b}_{\lambda} ) \tilde{s}
\nonumber\\
&& + v_0 \, \left[ ( b^{\dagger}_{\rho} \cdot b^{\dagger}_{\rho}
+ b^{\dagger}_{\lambda} \cdot b^{\dagger}_{\lambda} ) \tilde{s} \tilde{s}
+ s^{\dagger} s^{\dagger} ( \tilde{b}_{\rho} \cdot \tilde{b}_{\rho}
+ \tilde{b}_{\lambda} \cdot \tilde{b}_{\lambda} ) \right]
\nonumber\\
&& + \sum_{l=0,2} c_l \, \left[
( b^{\dagger}_{\rho} \times   b^{\dagger}_{\rho}
- b^{\dagger}_{\lambda} \times b^{\dagger}_{\lambda} )^{(l)} \cdot
( \tilde{b}_{\rho} \times \tilde{b}_{\rho}
- \tilde{b}_{\lambda} \times \tilde{b}_{\lambda} )^{(l)}
+ 4 \, ( b^{\dagger}_{\rho} \times b^{\dagger}_{\lambda})^{(l)} \cdot
       ( \tilde b_{\lambda} \times \tilde b_{\rho})^{(l)} \right]
\nonumber\\
&& + c_1 \, ( b^{\dagger}_{\rho} \times b^{\dagger}_{\lambda} )^{(1)}
\cdot ( \tilde b_{\lambda} \times \tilde b_{\rho} )^{(1)}
+ \sum_{l=0,2} w_l \, ( b^{\dagger}_{\rho} \times b^{\dagger}_{\rho}
  + b^{\dagger}_{\lambda} \times b^{\dagger}_{\lambda} )^{(l)} \cdot
  ( \tilde{b}_{\rho} \times \tilde{b}_{\rho}
  + \tilde{b}_{\lambda} \times \tilde{b}_{\lambda} )^{(l)} ~.
\nonumber\\ \label{ms3}
\ea
Here the dots indicate scalar products and the crosses tensor products
as usual. If the three objects are not identical (as is the case in
strange baryons) or if the interactions between the three objects are
not identical (as is the case if there is a flavor-spin dependence),
the mass-squared operator is no longer invariant under $S_3$, and a
more general form, still within $U(7)$, arises.
This situation will be discussed in a subsequent publication.

The eigenvalues and corresponding eigenvectors of the mass-squared
operator in Eq.~(\ref{ms3}) can be obtained exactly by
diagonalization in the basis of either Eq.~(\ref{basis1})
or~(\ref{basis2}). In Appendix~A we describe how the matrix
elements in either basis are calculated.
The wave functions obtained in this way have by construction good
angular momentum, parity and permutation symmetry. The permutation
symmetry of a given wave function is determined as follows. Firstly,
since Eq.~(\ref{ms3}) is invariant under the transposition $P(12)$,
basis states with $n_{\rho}$ even and $n_{\rho}$ odd do not mix, and
can therefore be treated separately. This allows one to distinguish
between states with $S,M_{\lambda}$ and states with $A,M_{\rho}$
symmetry. Secondly, we note that the operator $\hat K^2_y$
(where $\hat K_y$ is given in Eq.~(\ref{operators}))
commutes with the $S_3$--invariant mass-squared operator of
Eq.~(\ref{ms3}) and has expectation values, $K^2_y$ (with
$K_y=0,\pm 1,\pm 2,\ldots$).
Since the cyclic permutation $P(123)$ acts in Fock space as a rotation
of $\theta=2\pi/3$ induced by $\hat K_y$:
$\mbox{exp}[-i \, \theta \hat K_y]$, one finds
that for $|K_y|=0(\mbox{mod }3)$ the wave functions transform as
$S$ or $A$, whereas for $|K_y|=1,2(\mbox{mod }3)$ they transform as
$M_{\rho}$ or $M_{\lambda}$.
In conclusion, the transposition separates $S,M_{\lambda}$ from
$A,M_{\rho}$ and the cyclic permutation separates $S,A$ from
$M_{\rho},M_{\lambda}$. We note that the quantum number $K_y$ is
analogous to the label $m$ used in \cite{Hey}, since the operator
$\hat{n}_{\zeta}-\hat{n}_{\eta}$ of \cite{Hey} is equal to
$\hat{K}_y$.

Eq.~(\ref{ms3}) contains several models of baryon
structure. These models correspond to different choices of the
coefficients in Eq.~(\ref{ms3}). We mention in particular two classes of
models:

(i) Single-particle (harmonic oscillator) quark models. These
correspond to the choice $v_0=0$, {\it i.e.} no coupling between
different harmonic oscillator shells,
\ba
\hat M^2 &=& M^2_0 + \epsilon_s \, s^{\dagger}s
- \epsilon_p \, (b_{\rho}^{\dagger} \cdot \tilde{b}_{\rho}
+ b_{\lambda}^{\dagger} \cdot \tilde{b}_{\lambda}) +
\mbox{ anharmonic terms} ~,
\ea
or, introducing the number operators of Eq.~(\ref{operators})
\ba
\hat M^2 &=& M^2_0 + \epsilon_s \, \hat N + (\epsilon_p - \epsilon_s) \,
(\hat n_{\rho} + \hat n_{\lambda}) + \mbox{ anharmonic terms} ~.
\ea
The nonrelativistic harmonic oscillator quark
model \cite{IK} is a model of this type, although it is
written for the mass $\hat M$ rather than for $\hat M^2$,
\ba
\hat M &=& \frac{p_{\rho}^2}{2\mu} + \frac{p_{\lambda}^2}{2\mu}
+ \frac{3}{2} \kappa \rho^2 + \frac{3}{2}  \kappa \lambda^2
+ \mbox{ perturbations}
\nonumber\\
&=& \epsilon (-b_{\rho}^{\dagger} \cdot \tilde{b}_{\rho}
- b_{\lambda}^{\dagger} \cdot \tilde{b}_{\lambda}+ 3)
+ \mbox{ perturbations}
\nonumber\\
&=& \epsilon (\hat n_{\rho}+ \hat n_{\lambda}+3)
+ \mbox{ perturbations} ~,
\ea
with $\epsilon = \sqrt{3\kappa/\mu}$.
The perturbations involve both anharmonic terms and terms that
couple different shells, but the breaking is relatively small.
For example, the nucleon wave function in these type of models is
still dominated by the $n_{\rho}+n_{\lambda}=0$ component (typically
of the order $80 \%$ \cite{IKK}). The unperturbed harmonic oscillator
quark model corresponds algebraically to the
decomposition of $U(7)$ into $U(6) \otimes U(1)$. All results of these
models are contained in $U(7)$, provided that the expansion in terms
of elements $G_{\alpha \alpha^{\prime}}$ is made for the mass operator
rather than its square. In Figure~\ref{harmosc} we show the spectrum for
the unperturbed harmonic oscillator.

(ii) Collective (string) models. In these models the three constituent
parts move in a correlated way. They correspond to the choice
$v_0 \neq 0$. Since in this case the mass-squared operator contains
terms of the type $b^{\dagger}b^{\dagger}ss+s^{\dagger}s^{\dagger}bb$,
the corresponding eigenfunctions are spread over many oscillator
shells. In order to study models of this type (which is the main and
novel purpose of this paper), it is instructive to rewrite the mass-squared
operator of Eq.~(\ref{ms3}) in terms of vibrational and rotational
contributions to the mass spectrum.
The general procedure for such a decomposition was introduced \cite{KL}
for the Interacting Boson Model in nuclear physics. Here we apply
the same method to the $S_3$-invariant mass operator of Eq.~(\ref{ms3})
\cite{BL}
\ba
\hat M^2 &=& \hat M^2_0 + \hat M^2_{\mbox{vib}}
+ \hat M^2_{\mbox{rot}} + \hat M^2_{\mbox{vib--rot}} ~, \label{mrv}
\ea
with
\ba
\hat M^{2}_{\mbox{vib}} &=& \xi_1 \,
\Bigl ( R^2 \, s^{\dagger} s^{\dagger}
- b^{\dagger}_{\rho} \cdot b^{\dagger}_{\rho}
- b^{\dagger}_{\lambda} \cdot b^{\dagger}_{\lambda} \Bigr ) \,
\Bigl ( R^2 \, \tilde{s} \tilde{s} - \tilde{b}_{\rho} \cdot \tilde{b}_{\rho}
- \tilde{b}_{\lambda} \cdot \tilde{b}_{\lambda} \Bigr )
\nonumber\\
&& + \xi_2 \, \Bigl [
\Bigl( b^{\dagger}_{\rho} \cdot b^{\dagger}_{\rho}
- b^{\dagger}_{\lambda} \cdot b^{\dagger}_{\lambda} \Bigr ) \,
\Bigl ( \tilde{b}_{\rho} \cdot \tilde{b}_{\rho}
- \tilde{b}_{\lambda} \cdot \tilde{b}_{\lambda} \Bigr )
+ 4 \, \Bigl ( b^{\dagger}_{\rho} \cdot b^{\dagger}_{\lambda} \Bigr ) \,
\Bigl ( \tilde{b}_{\lambda} \cdot \tilde{b}_{\rho} \Bigr ) \Bigr ] ~,
\nonumber\\
\hat M^{2}_{\mbox{rot}} &=&
  2 \xi_3 \, \hat G_S^{(1)} \cdot \hat G_S^{(1)}
+ 3 \xi_4 \, \hat G_A^{(0)} \cdot \hat G_A^{(0)}
\nonumber\\
&=& \xi_3 \, \hat L \cdot \hat L + \xi_4 \, \hat K_y \cdot \hat K_y ~,
\nonumber\\
\hat M^{2}_{\mbox{vib--rot}} &=& \xi_5 \, \left[
\hat A_{\rho} \cdot \hat A_{\rho} + \hat A_{\lambda} \cdot
\hat A_{\lambda} \right]
+ \xi_6 \, \left[ \hat G^{(1)}_{M_{\rho}} \cdot \hat G^{(1)}_{M_{\rho}}
+ \hat G^{(1)}_{M_{\lambda}} \cdot \hat G^{(1)}_{M_{\lambda}} \right] ~,
\nonumber\\
\hat M^2_0 &=& \xi_7 + \xi_8 \, (\hat n_{\rho} + \hat n_{\lambda}) ~.
\label{mrv1}
\ea
The explicit expression of $\hat G^{(1)}_S,\ldots,$ in terms of boson
operators is given in Eqs.~(\ref{gen}) and~(\ref{operators}).
By rewriting $\hat M^2$ in this form one emphasizes the physical content
of the string-like model, since the excitation spectrum will now appear
as vibrations and rotations of the string-like configuration shown in
Figure~\ref{geometry}. The new parameters in Eq.~(\ref{mrv1}) are linear
combinations of those in Eq.~(\ref{ms3}).

Although the mass spectrum and corresponding eigenfunctions of $\hat M^2$
can be obtained numerically by diagonalization, we prefer here to use
coherent states to gain further insight into the physical content of
each contribution. To this end, we use the coherent (or intrinsic) states of
Appendix~B. We begin by considering the vibrational term
$\hat M^{2}_{\mbox{vib}}$. The Bose condensate of Eq.~(\ref{bc}) is the
lowest eigenstate of this term ($\xi_1$, $\xi_2>0$).
Indeed the separation (\ref{mrv}) has been done in order to satisfy
this condition. For large $N$ higher eigenstates are of the type
(\ref{cuvw}) and, to leading order in $N$, the corresponding eigenvalues are
given by \cite{BL}
\ba
M^2_{\mbox{vib}} &=& N \left[ \kappa_1 \, n_u
+ \kappa_2 \, (n_v+n_w) \right] ~,
\ea
where $n_u,n_v,n_w \, (\geq 0)$ are the eigenvalues of the number operators
$b^{\dagger}_u b_u$, $b^{\dagger}_v b_v$, $b^{\dagger}_w b_w$
(the $b_u^{\dagger}$, $b_v^{\dagger}$, $b_w^{\dagger}$ operators are
given in Eq.~(\ref{bvib})), and
\ba
\kappa_1 &=& 4 \, \xi_1 \, R^2 ~,
\nonumber\\
\kappa_2 &=& 4 \, \xi_2 \, R^2 (1+R^2)^{-1} ~. \label{evib}
\ea
The vibrational part of the mass-squared operator of Eq.~(\ref{mrv})
has a very simple physical interpretation. Its spectrum has three
fundamental vibrations (see Figure~\ref{vibrations}).
The $u$--vibration is the
symmetric stretching vibration along the direction of the strings
(breathing mode), while the $v$-- and the $w$--vibrations denote
bending vibrations of the strings. The latter two vibrations are
degenerate in the case of three identical objects \cite{Herzberg}.
We note in passing that QCD based arguments suggest that
while the string is soft towards stretching, it is hard towards bending
and thus one expects the $v$-- and $w$--vibrations to lie higher than the
$u$--vibration.

Next we discuss the rotational part of the mass-squared operator.
It contains two terms that commute with the general $S_3$--invariant mass
operator of Eq.~(\ref{ms3}) and hence correspond to exact symmetries.
The eigenvalues of these rotational terms are
\ba
M^2_{\mbox{rot}} &=& \xi_3 \, L(L+1) + \xi_4 \, K^2_y ~.
\label{erot}
\ea
Here $L$ is the orbital angular momentum and $K_y$ corresponds in
the large $N$ limit to the projection $K$ of the angular momentum on
the threefold symmetry axis (the $y$--axis in Figure~\ref{variables}).
For the ground state band the values of $L$ and $K_y$ are
\ba
L &=& 0,1,2,\ldots,L_{\mbox{max}} ~,
\nonumber\\
K_y &=& 0,\pm 1, \cdots, \pm L ~.
\ea
For the excited bands the situation is slightly more complicated
and it will not be discussed here.
Again the physical interpretation of Eq.~(\ref{erot}) is simple,
since it describes the rotational spectrum of the string-like
configuration of Figure~\ref{geometry}. For finite $N$ the rotational
spectrum is truncated at a finite value of $L=L_{\mbox{max}}$, while
for $N \rightarrow \infty$ also $L_{\mbox{max}} \rightarrow \infty$.

The discussion up to this point applies to any object with the
geometric configuration of Figure~\ref{geometry}.
However, the rotational spectrum
of Eq.~(\ref{erot}) does not reproduce a characteristic feature of
hadronic spectra (expected on the basis of QCD \cite{JT} and
investigated decades ago), namely, the occurrence of linear Regge
trajectories. But, since $L$ and $K_y$ are good quantum numbers
one can consider, still remaining within $U(7)$, more complicated
functional forms, $f(\hat L^2)+g(\hat K_y^2)$, with eigenvalues,
$f(L(L+1))+g(K_y^2)$. Linear Regge trajectories are simply obtained
by choosing the form
\ba
\hat M^2_{\mbox{rot}} &=& \alpha \, \sqrt{\hat L \cdot \hat L + 1/4}
+ \beta \, \sqrt{\hat K_y \cdot \hat K_y} ~,
\ea
with eigenvalues
\ba
M^2_{\mbox{rot}} &=& \alpha \, (L+1/2) + \beta \, |K_y| ~. \label{frot}
\ea
The rotational spectrum of Eq.~(\ref{erot}) and~(\ref{frot}) is that
of an oblate top \cite{Herzberg}.
If $\xi_4=0$ (or $\beta=0$) the top is symmetric. When viewed as a
top, the configuration of Figure~\ref{geometry} has $D_3$ point group
symmetry. Since $D_3$ is isomorphic to $S_3$, one can
label the states either with $S$, $M$ and $A$, or with the equivalent
labels of $D_3$, {\it i.e.} $A_1$, $E$ and $A_2$,
\ba
S \leftrightarrow A_1 ~, \hspace{1cm} M \leftrightarrow E ~,
\hspace{1cm} A \leftrightarrow A_2 ~.
\ea

Unlike the rotational part, the last term in the $S_3$--invariant
mass-squared operator of Eq.~(\ref{mrv}) does not commute with the
vibrational part and hence introduces vibration-rotation interactions.
We shall not discuss this term any further, since the experimental
mass spectrum of baryons is not known accurately enough to be able
to detect rotation-vibration couplings.
Finally, the $\hat M^2_0$ term in Eq.~(\ref{mrv}) contains
an overall constant that does not contribute to mass splittings
and a one-body term whose contribution is negligible in the
large $N$ limit.

We summarize the results of the present analysis by showing in
Figure~\ref{string} the spectrum of the string-like configuration of
Figure~\ref{geometry}, given by the simplified mass formula
\ba
M^2 &=& M^2_0 + N \left[ \kappa_1 \, n_u
+ \kappa_{2} \, (n_v+n_w) \right] + \alpha \, L ~. \label{mspace}
\ea
Here we have discarded the $K_y$--dependent term in the rotational
part, since the experimental mass spectrum of nonstrange baryons does not
provide any compelling evidence for its presence. In Figure~\ref{string}
we have used the projection $K$ of the angular momentum (equal to the
algebraic $K_y$ for the ground and first excited vibrational band).
All constant contributions have been absorbed into $M^2_0$.
A comparison with the mass spectrum of the harmonic oscillator
in Figure~\ref{harmosc} shows that whereas for the harmonic oscillator
the excited $L^{\pi}=0^+$ states belong to the two-phonon
($n=n_{\rho}+n_{\lambda}=2$) multiplet, in the collective string model
they correspond to one-phonon vibrational excitations and are the
bandheads of these fundamental vibrations.

\subsection{Spin-flavor part}

In the previous subsection we have discussed the space part of the
mass-squared operator with $S_3$ symmetry.
We turn now to a discussion of the internal degrees of freedom of
the constituent parts and construct the corresponding contribution to
the mass-squared operator and their eigenfunctions.
The spatial part of the baryon wave function, which is determined by
the $U(7)$ mass-squared operator, has to be combined with the spin-flavor
and color part, in such a way that the total wave function is antisymmetric
\ba
| \psi \rangle &=& | \psi_L \rangle \otimes | \psi_{sf} \rangle
\otimes | \psi_{c} \rangle ~.
\ea
Here $|\psi_L \rangle$ denotes the space part, $|\psi_{sf} \rangle$
the spin-flavor part and $|\psi_{c} \rangle$ the color part.
Since the color part
of the wave function is totally antisymmetric (color singlet),
the remaining part (space plus spin-flavor) must be totally symmetric.
This implies, for the case of three identical constituent parts
(discussed here), that the symmetry of $|\psi_L \rangle$ under $S_3$ is
the same as the symmetry of $|\psi_{sf} \rangle$.
The construction of spin-flavor wave functions with good $S_3$
symmetry is well-known (see, for example, Refs.~\cite{IK,FKR,KI}) and
in Appendix~C we list the conventions used. We denote the basis states
for the spin-flavor part by
\ba
\left| \begin{array}{ccccccccccc}
SU_{sf}(6) &\supset& SU_{f}(3) &\otimes& SU_{s}(2)
&\supset& SU_{I}(2) &\otimes& U_{Y}(1) &\otimes&
SO_{s}(2) \\
\downarrow && \downarrow && \downarrow && \downarrow && \downarrow &&
\downarrow \\
\, [f_1,f_2,\ldots,f_5] && [\mu_1,\mu_2] && S && I && Y && M_S
\end{array} \right> ~. \label{sfbasis}
\ea
Here $S$ denotes the spin, $I$ the isospin and $Y$ the hypercharge.
An unfortunate (but standard) notation is to label representations
not by their Young tableaux but by their dimension. For example,
for $SU_{f}(3)$,
\ba
\mbox{dim}[\mu_1,\mu_2] &=&
\frac{1}{2} (\mu_1+2)(\mu_2+1)(\mu_1-\mu_2+1) ~,
\ea
which gives 8 for $[\mu_1,\mu_2]=[2,1]$ and 10 for [3,0]. In the following
sections we adopt the standard notation in order to facilitate the
comparison with other model calculations. Thus, for example,
\ba
\left| [3,0,0,0,0],[2,1],S=\frac{1}{2},I=\frac{1}{2},Y=1 \right>
&\equiv& | [56],^{2}8,\mbox{N} \rangle ~, \label{nucleon}
\ea
represents the spin-flavor part of the wave function of the ground
state of the nucleon. The decomposition of representations of
$SU_{sf}(6)$ into those of $SU_{f}(3) \otimes SU_{s}(2)$
is the standard one
\ba
\, [56] &\supset& ^{2}8 \, \oplus \, ^{4}10 ~,
\nonumber\\
\, [70] &\supset& ^{2}8 \, \oplus \, ^{4}8 \, \oplus \, ^{2}10 \,
\oplus \, ^{2}1 ~,
\nonumber\\
\, [20] &\supset& ^{2}8 \, \oplus \, ^{4}1 ~.
\ea

The spin-flavor contribution to the mass-squared operator can be
expressed in terms of the generators of the spin-flavor algebra.
We consider here only its diagonal part which we write in the
G\"ursey-Radicati \cite{GR} form
\ba
\hat M^2_{sf}
&=& a \, \Bigl[ \hat C_2(SU_{sf}(6)) - 45 \Bigr]
+ b \, \Bigl[ \hat C_2(SU_{f}(3)) -  9 \Bigr]
+ b^{\prime} \, \Bigl[ \hat C_2(SU_{I}(2)) -  \frac{3}{4} \Bigr]
\nonumber\\
&& + b^{\prime\prime}       \, \Bigl[ \hat C_1(U_{Y}(1)) -1 \Bigr]
   + b^{\prime\prime\prime} \, \Bigl[ \hat C_2(U_{Y}(1)) -1 \Bigr]
+ c \, \Bigl[ \hat C_2(SU_{s}(2)) - \frac{3}{4} \Bigr] ~.
\ea
We have defined the operators such that each of the terms vanishes
for the ground state of the nucleon (see Eq.~({\ref{nucleon})). The
eigenvalues of the Casimir operators in the basis states of
Eq.~(\ref{sfbasis}) are
\ba
\langle \hat C_2(SU_{sf}(6)) \rangle &=& \left\{ \begin{array}{cc}
45 & \mbox{ for } 56 \leftrightarrow A_1 \leftrightarrow S \\
33 & \mbox{ for } 70 \leftrightarrow E \leftrightarrow M \\
21 & \mbox{ for } 20 \leftrightarrow A_2 \leftrightarrow A \end{array}
\right. ~,
\nonumber\\
\langle \hat C_2(SU_{f}(3)) \rangle &=& \left\{ \begin{array}{cc}
9  & \mbox{ for } 8 \\ 18 & \mbox{ for } 10 \\
0  & \mbox{ for } 1 \end{array} \right. ~,
\nonumber\\
\langle \hat C_2(SU_{I}(2)) \rangle &=& I(I+1) ~,
\nonumber\\
\langle \hat C_1(U_{Y}(1)) \rangle &=& Y ~,
\nonumber\\
\langle \hat C_2(U_{Y}(1)) \rangle &=& Y^2 ~,
\nonumber\\
\langle \hat C_2(SU_{s}(2)) \rangle &=& S(S+1) ~. \label{sfcasimir}
\ea
For nonstrange baryons $Y=1$ and the $b^{\prime\prime}$ and
$b^{\prime\prime\prime}$ terms are not needed. Also the $b$ and
$b^{\prime}$ terms can be grouped into a single term. Thus, for the
analysis of nonstrange baryons we make use of a simplified form
\ba
\hat M^2_{sf} &=&
  a \, \Bigl[ \hat C_2(SU_{sf}(6)) - 45 \Bigr]
+ b \, \Bigl[ \hat C_2(SU_{f}(3)) -  9 \Bigr]
+ c \, \Bigl[ \hat C_2(SU_{s}(2)) - \frac{3}{4} \Bigr] ~.
\label{msf}
\ea
with three parameters. The meaning of the three terms is obvious.
The spin term represents spin-spin interactions, the flavor term
denotes the flavor dependence of the interactions, and the
$SU_{sf}(6)$ term, which according to Eq.~(\ref{sfcasimir})
depends on the permutation symmetry of the wave functions, represents
`signature dependent' interactions. These signature dependent (or
exchange) interactions were extensively investigated years ago within
the framework of Regge theory \cite{exchange}.
We note in passing that in the usual
nonrelativistic and relativized quark models the spin-flavor dependence
arises from (i) the constituent quark masses producing a dependence
similar to that of the $b^{\prime\prime}$ term and (ii) from spin-spin
forces $\vec{\sigma}_i \cdot \vec{\sigma}_j$, that arise from the contact
term of the hyperfine interaction, producing a dependence of the type
$cS(S+1)$. In this sense, Eq.~(\ref{msf}) is more
general than the corresponding spin-flavor dependence in the quark model.

Finally, there could be terms in the mass-squared operator involving
simultaneously both internal and spatial degrees of freedom, {\it i.e.}
of the type
\ba
\hat M^2 &=& f(G_{\alpha \alpha^{\prime}}) \, g(G_i) ~,
\hspace{1.cm} G_{\alpha \alpha^{\prime}} \; \in \; {\cal G}_r ~,
\hspace{.5cm} G_i \; \in \; {\cal G}_i ~.
\ea
Among these, we mention: (i) spin-orbit-like interactions
\ba
\hat M^2_{\mbox{so}} &=& d \, (\vec{S} \cdot \vec{L}) ~,
\ea
and (ii) tensor-like interactions
\ba
\hat M^2_{\mbox{tensor}} &=&
d^{\prime} \, (\hat T_S^{(2)} \cdot \hat T_L^{(2)}) ~, \label{tensor}
\ea
where $\hat T_S^{(2)}$ and $\hat T_L^{(2)}$ are tensors of rank 2 built
from the spin and space degrees of freedom. Both terms are present in the
nonrelativistic and relativized quark models, in which the $SU_{sf}(6)$
spin-flavor symmetry is broken by the hyperfine interaction
(Eq.~(\ref{tensor}) corresponds to the tensor component
of the hyperfine interaction). Although in the present analysis we
do not consider the contributions of the spin-orbit and the tensor
interactions, we note that in the algebraic approach
they can be taken into account without any difficulty.

\section{Comparison with experimental mass spectrum}
\setcounter{equation}{0}

The mass formulas derived in Section~5 can be used to analyze the
experimental mass spectrum of baryons. Here we discuss the mass spectrum
of the nonstrange baryons belonging to the N and $\Delta$ families
in terms of the mass formula
\ba
M^2 &=& M^2_0 + (\kappa_1 N) \, n_u + (\kappa_2 N) \, (n_v+n_w)
+ \alpha \, L
\nonumber\\ &&
+ a \, \Bigl[ \langle \hat C_2(SU_{sf}(6)) \rangle - 45 \Bigr]
+ b \, \Bigl[ \langle \hat C_2(SU_{f}(3)) \rangle -  9 \Bigr]
+ c \, \Bigl[ \langle \hat C_2(SU_{s}(2)) \rangle
- \frac{3}{4} \Bigr] ~. \label{massformula}
\ea
In this form we have absorbed all constant terms into $M^2_0$. We do
not include interaction terms that mix the space and internal degrees
of freedom. The seven coefficients are obtained by a fit to the
data
\ba
M_{0}^{2}=0.882 ~, \hspace{1cm} \kappa_1 N=1.192 ~,
\hspace{1cm} \kappa_2 N=1.535 ~, \hspace{1cm}
\nonumber\\
\alpha=1.064 ~, \hspace{1cm} a=-0.042 ~, \hspace{1cm} b=0.030 ~,
\hspace{1cm} c=0.124 ~.
\ea
All values are in GeV$^2$. The results are given in Table~\ref{masses}.
We have assigned the Roper resonance N$(1440)P_{11}$, the
$\Delta(1600)P_{33}$ and the $\Delta(1900)S_{31}$ to the
symmetric stretching vibration $(n_u,n_v+n_w)=(1,0)$ and the
resonance N$(1710)P_{11}$ to the $(n_u,n_v+n_w)=(0,1)$ vibration.
The remaining part of the assignments is straightforward as rotational
members of the ground band $(n_u,n_v+n_w)=(0,0)$.
In Table~\ref{masses} we list all well-established (3 and 4 star)
nucleon and delta resonances. We find a good overall fit for these
resonances with an r.m.s. deviation of $\delta_{\mbox{rms}}=39$ MeV.
Some of the results are also presented in Figure~\ref{regge} in a
standard Chew-Frautschi plot of baryon resonances \cite{Chew}.

It is worthwhile at this stage to comment on the results of
Table~\ref{masses}. We find that the spin-orbit interaction is not
required by the data. In the quark model, where this term
is introduced by the one-gluon exchange interaction, other mechanisms
have to be found to cancel this contribution (called the `spin-orbit
crisis'). We find instead that the `signature dependent' terms
are crucial in obtaining a good description of the data. In the
quark model, this problem is solved either by using harmonic
oscillator frequencies which are different for P-wave and D-wave
baryons \cite{IK}, or by giving only a qualitative description
which is somewhat lower for P-wave and somewhat higher for D-wave
baryons \cite{CI}. This situation is illustrated in Figure~\ref{signature}.
The absence of the spin-orbit interaction is evident in the data
(states with the same $L$, $S$ and $|L-S| \leq J \leq L+S$
have the same mass). For comparison, in Figure~\ref{signature} we also
show the results of \cite{CI}.
Finally the value we find for $\alpha$, the slope of the Regge trajectory,
is almost identical to that found in mesons \cite{meson}
\ba
\alpha_{\mbox{meson}} &=& 1.081 \mbox{ GeV}^2 ~,
\nonumber\\
\alpha_{\mbox{baryon}} &=& 1.064 \mbox{ GeV}^2 ~.
\ea
This is consistent with QCD ideas which indicate a universal slope
for both baryons and mesons \cite{JT}. The strength of the spin-spin
interaction is also almost identical to that found in mesons:
$c_{\mbox{meson}} =0.118 \mbox{ GeV}^2$ and
$c_{\mbox{baryon}}=0.124 \mbox{ GeV}^2$.

For completeness, we give in Tables~\ref{missingn} and~\ref{missingd}
all calculated nucleon and delta resonances below 2 GeV.
The lowest missing states in the nucleon sector are the antisymmetric
$[20,1^+]$ ones. As one can see from these tables, only a fraction of
these resonances have been observed. This problem of missing states is
known to exist in quark potential models as well.
The resonances in square brackets are not well established experimentally
(1 and 2 star) and are tentatively assigned in the tables as candidates
for some of the missing states. We shall
return to the question of why some of the resonances have not been
observed in a later publication, which includes a calculation of the
strong decay widths.

To summarize, we have applied a collective string-like model to the
nonstrange baryon resonances and found good overall agreement with
the observed masses. The fit is of comparable quality to that obtained
in quark potential models \cite{IK,CI}, although the underlying dynamics
is quite different. This shows that masses alone are not sufficient to
distinguish between different forms of quark dynamics, {\it e.g.}
single-particle {\it vs.} collective motion.

\section{Electromagnetic couplings}
\setcounter{equation}{0}

Electromagnetic couplings are of crucial importance in unraveling
the structure of hadrons, since they are far more sensitive to wave
functions (and models) than masses. It has become customary to
characterize the transverse couplings by the helicity amplitudes,
$A_{1/2}$ and $A_{3/2}$. These amplitudes are measurable
in photo- and electroproduction. Their study is a major part of the
experimental program at the new electron facilities \cite{Burkert}.

Helicity amplitudes can be computed within the framework discussed
here by (i) writing down the transition operator in terms of the
constituent coordinates and momenta, (ii) rewriting them in terms
of Jacobi coordinates and momenta, (iii) mapping the operators onto
elements of the algebra and (iv) evaluating their matrix elements
algebraically.
A major problem in step (i) is what is precisely the form of the
electromagnetic coupling. This form is usually obtained by a
nonrelativistic reduction of the coupling of the point-like
particles to the electromagnetic field. This gives a transition
operator of the form \cite{CL,Warns}
\ba
{\cal H} &=& {\cal H}_{\mbox{nr}} + {\cal H}_{\mbox{so}}
+ {\cal H}_{\mbox{na}} + \ldots ~,
\ea
containing the contributions of the nonrelativistic part,
the spin-orbit part and the nonadditive part associated with
Wigner rotations \cite{CL} and higher order corrections,
\ba
{\cal H}_{\mbox{nr}} &=& - \sum_{j=1}^{3} \left[ \frac{e_j}{2m_j}
(\vec{p}_j \cdot \vec{A}_j + \vec{A}_j \cdot \vec{p}_j)
+ 2\mu_j \vec{s}_j \cdot (\vec{\nabla} \times \vec{A}_j) \right] ~,
\nonumber\\
{\cal H}_{\mbox{so}} &=& - \sum_{j=1}^{3} \mu_j \frac{1}{2m_j}
(2-\frac{1}{g}) \, \vec{s}_j \cdot
(\vec{E}_j \times \vec{p}_j - \vec{p}_j \times \vec{E}_j) ~,
\nonumber\\
{\cal H}_{\mbox{na}} &=& \frac{1}{2M_T} \sum_{j>i=1}^{3}
( \frac{\vec{s}_i}{m_i} - \frac{\vec{s}_j}{m_j} ) \cdot
( e_j \vec{E}_j \times \vec{p}_i - e_i \vec{E}_i \times \vec{p}_j ) ~.
\label{hem}
\ea
where $m_j$, $e_j$, $\vec{s}_j$ and $\mu_j=g e_j/2m_j$ denote the mass,
charge, spin and magnetic moment of the $j$--th constituent, respectively,
$M_T=\sum_i m_i$ and $\vec{A}_j \equiv \vec{A}(\vec{r}_j)$,
$\vec{E}_j \equiv \vec{E}(\vec{r}_j)$.
For purposes of illustrating the results in a transparent way, we
consider in this article only the contribution from the nonrelativistic
part of the electromagnetic coupling,
\ba
{\cal H} &=& {\cal H}_{\mbox{nr}} ~.
\ea
The spin-orbit and non-additive contributions can be included without
any problem and will be presented in a subsequent publication.

The momentum dependent terms in Eq.~(\ref{hem}) are unsuited for
calculations of electromagnetic couplings of the photon to an extended
object, in which the charge and magnetic moment are not concentrated at a
single point but distributed along the string. Thus, often a transformation
is made to coordinate dependent terms by replacing $\vec{p}/m_q$ by
$i k_0 \vec{r}$ \cite{CB,FD}, where $k_0=E_f-E_i$ is the photon energy. The
two terms in ${\cal H}_{\mbox{nr}}$ have then the meaning of electric and
magnetic contributions and reduce to the electric dipole and magnetic dipole
transition operators in the long-wavelength limit.
The transverse coupling is obtained by inserting the radiation
field for the absorption of a righthanded photon with momentum
$\vec{k}=k \hat z$ and integrating over the
center-of-mass coordinate to give
\ba
{\cal H}^t &=& 6 \sqrt{\frac{\pi}{k_0}} \mu e_3 \,
\left[ k s_{3,+} \hat U - \frac{1}{g} \hat T_{+} \right] ~,
\ea
Here ($k_0,\vec{k}$) is the photon four-momentum with
$\vec{k}=\vec{P}_f-\vec{P}_i$, and $\vec{P}_i$ ($\vec{P}_f$) is
the momentum of the initial (final) baryon resonance. In the derivation we
have used that for three identical constituents the symmetry of the
wave functions allows one to write ${\cal H}=3{\cal H}_3$.
The operators $\hat T$ and $\hat U$ act only on the spatial part
of the baryon wave function and are given by
\ba
\hat U &=& \mbox{e}^{ -ik \sqrt{\frac{2}{3}} \lambda_z} ~,
\nonumber\\
\hat T_{m} &=& i m_q k_0 \sqrt{\frac{2}{3}} \, \lambda_m \,
\mbox{e}^{ -ik \sqrt{\frac{2}{3}} \lambda_z} ~,
\ea
with $m=-1,0,1$.

In addition to transverse couplings, one can also consider longitudinal
and scalar couplings. The longitudinal coupling is obtained by inserting
the radiation field for the absorption of a longitudinally polarized
virtual photon in Eq.~(\ref{hem}),
\ba
{\cal H}^l &=& 6 \sqrt{\frac{2\pi}{k_0}} \mu e_3 \,
\frac{1}{g} \, \hat T_z ~.
\ea

The scalar amplitudes are given by the matrix elements of
the zero-component of the electromagnetic current four-vector
\cite{BM}
\ba
{\cal H}^s &=& - 3 \sqrt{\frac{2\pi}{k_0}} \, e_3 \,
\mbox{e}^{ ik r_{3,z} } ~,
\ea
which after transforming to Jacobi coordinates and integrating
over the baryon center-of-mass coordinate reduces to
\ba
{\cal H}^s &=& - 3 \sqrt{\frac{2\pi}{k_0}} \, e_3 \,
\hat U ~.
\ea

In order to calculate helicity amplitudes in $U(7)$ one has
to express the transition operators in terms of algebraic operators.
In the limit of large $N$ the operators $\hat D$ and $\hat A$ of
Eq.~(\ref{gen}) become the coordinates and momenta ({\it i.e.} their
matrix elements are precisely those of the coordinates and
momenta in the coordinate representation). We thus make in general
the replacement (mapping) \cite{onno}
\ba
\sqrt{\frac{2}{3}} \, \lambda_m &\rightarrow&
\beta \, \hat D_{\lambda,m}/X_D ~,
\nonumber\\
\sqrt{\frac{2}{3}} \, p_{\lambda,m} &\rightarrow&
\frac{1}{\zeta} \, \hat A_{\lambda,m}/X_A ~, \label{map}
\ea
and a similar replacement for $\rho_m$, $p_{\rho,m}$ (not
needed here). In Eq.~(\ref{map}), $\beta$ and $1/\zeta$ represent the
scale of coordinates and momenta and $X_D$ and $X_A$ are
normalization factors.
Since we have written the transition operators in terms of
coordinates only, we need only to replace $\lambda_m$ by
$\hat D_{\lambda,m}$. With this replacement the transition
operators $\hat T_m$ and $\hat U$ become
\ba
\hat U &=& \mbox{e}^{ -i k \beta \hat D_{\lambda,z}/X_D } ~,
\nonumber\\
\hat T_m &=& \frac{i m_q k_0 \beta}{2 X_D} \left( \hat D_{\lambda,m} \,
\mbox{e}^{ -i k \beta \hat D_{\lambda,z}/X_D } +
\mbox{e}^{ -i k \beta \hat D_{\lambda,z}/X_D } \, \hat D_{\lambda,m}
\right) ~. \label{emop}
\ea
The normalization factor is given by the reduced matrix element of
$\hat D_{\lambda}$,
\ba
X_D &=& |\langle 1^-_{\lambda} || \hat D_{\lambda} || 0^+_S \rangle| ~,
\ea
which away from the harmonic oscillator limit ($R^2 > 0$) is given by
\ba
X_D &\stackrel{N \rightarrow \infty}{\longrightarrow}&
NR\sqrt{2}/(1+R^2) ~.
\ea
In the harmonic oscillator limit discussed in Section~5
the normalization constant is $X_D=\sqrt{3N}$ and
the scale parameter $\beta$ becomes the inverse of the
oscillator size parameter $\beta=1/\alpha$ (see Appendix~D).

The algebraic structure of the transverse, longitudinal and scalar
couplings involves both the internal
and the spatial degrees of freedom. In the long-wavelength limit
the spatial part is linear in the generators of $U(7)$, but the
more general form of Eq.~(\ref{emop}) also contains an exponentiated
generator. This poses a challenge to the calculation. Nonetheless, the
calculation of the matrix elements of Eq.~(\ref{emop}) is feasible,
since they are related to the group elements of $U(7)$,
{\it i.e} a generalization of the familiar Wigner D-functions
of $SU(2)$. This means that these matrix elements can be
calculated exactly without having to make any further approximation.
This holds for the harmonic oscillator which is a special case
of $U(7)$ as well,
and thus allows one to do a straightforward calculation of helicity
amplitudes in the quark model up to large number of quanta. Indeed,
the fact that any observable can be calculated in a relatively
straightforward way, is one of the main advantages of the
algebraic method.

\section{Calculation of helicity amplitudes}
\setcounter{equation}{0}

The calculation of the helicity amplitudes requires the evaluation
of the matrix elements of the electromagnetic transition operator.
We define the transverse helicity amplitudes $A_{\mu}$, with helicity
$\mu=1/2$ and 3/2 in the usual fashion
\ba
A_{1/2} &=&
\langle \phi^{\prime},L^{\prime},S^{\prime};J^{\prime},
M_J^{\prime}=1/2| {\cal H}^t |\phi,L,S;J=1/2,M_J=-1/2 \rangle ~,
\nonumber\\
A_{3/2} &=&
\langle \phi^{\prime},L^{\prime},S^{\prime};J^{\prime},
M_J^{\prime}=3/2| {\cal H}^t |\phi,L,S;J=1/2,M_J= 1/2 \rangle ~.
\ea
The longitudinal and scalar helicity amplitudes, $A_l$ and
$A_s$, are given by
\ba
A_l &=&
\langle \phi^{\prime},L^{\prime},S^{\prime};J^{\prime},
M_J^{\prime}=1/2| {\cal H}^l |\phi,L,S;J=1/2,M_J=1/2 \rangle/\sqrt{2} ~,
\nonumber\\
A_s &=&
\langle \phi^{\prime},L^{\prime},S^{\prime};J^{\prime},
M_J^{\prime}=1/2| {\cal H}^s |\phi,L,S;J=1/2,M_J=1/2 \rangle/\sqrt{2} ~.
\ea
Here $\phi$ indicates all additional quantum numbers needed to
classify the states uniquely.
The electromagnetic transition operator ${\cal H}$ acts both on
the spin-flavor part and the space part of the
baryon wave functions. In the evaluation of the matrix elements we
therefore separate these two parts by decoupling the wave functions
\ba
|\phi,L,S;J,M_J \rangle &=& \sum_{M_L,M_S}
\langle L,M_L,S,M_S|J,M_J \rangle \, |\phi,L,M_L;S,M_S \rangle ~.
\ea
The spin-flavor part of the matrix elements is common to all models
having the same spin-flavor structure and can be evaluated once and
for all.

In general the helicity amplitudes can be expressed explicitly
in terms of radial integrals. For the transverse couplings we
have
\ba
A_{\mu} \;=\; \langle f|{\cal H}^t|i \rangle \;=\;
\alpha_{\mu} \, {\cal A} + \beta_{\mu} \, {\cal B} ~, \label{helampt}
\ea
where $\mu$ denotes the helicity ($\mu=1/2,3/2$). Here ${\cal A}$ and
${\cal B}$ represent the orbit- and spin-flip spatial amplitudes
(radial integrals), respectively,
\ba
{\cal A} &=& 6 \sqrt{\frac{\pi}{k_0}} \, \mu \frac{1}{g} \,
\langle f | \hat T_+ | i \rangle ~,
\nonumber\\
{\cal B} &=& 6 \sqrt{\frac{\pi}{k_0}} \, \mu k \,
\langle f | \hat U | i \rangle ~.
\ea
The nonrelativistic contribution to the longitudinal helicity
amplitudes can be expressed in terms of a single spatial amplitude
\ba
A_l \;=\: \langle f|{\cal H}^l|i \rangle/\sqrt{2}
\;=\; \gamma \, {\cal C} ~, \label{helampl}
\ea
with
\ba
{\cal C} &=& 6 \sqrt{\frac{\pi}{k_0}} \mu \frac{1}{g} \,
\langle f | \hat T_{z} | i \rangle ~.
\ea
For the scalar helicity amplitudes we find
\ba
A_s \;=\: \langle f|{\cal H}^s|i \rangle/\sqrt{2}
\;=\; \delta \, {\cal D} ~, \label{helamps}
\ea
with
\ba
{\cal D} &=& - 3 \sqrt{\frac{\pi}{k_0}} \,
\langle f | \hat U | i \rangle ~.
\ea

The coefficients $\alpha_{\mu}$, $\beta_{\mu}$, $\gamma$ and
$\delta$  contain the contribution of the spin-flavor matrix element
and of Clebsch-Gordan coefficients. By explicitly evaluating the matrix
elements of the spin-flavor part with the conventions given in
Appendix~C, we obtain the coefficients for nucleon and delta resonances
as shown in Tables~\ref{emn} and~\ref{emd}. Some of these
results have already been reported previously (see {\it e.g.}
\cite{CL,SS}) and express the fact that all models of hadronic structure,
that share the same spin-flavor structure, have the same form of the
helicity amplitudes. The model dependence is contained in the actual
expression for the spatial amplitudes, ${\cal A}$, ${\cal B}$, ${\cal C}$
and ${\cal D}$. We therefore now turn to the calculation of these amplitudes
in $U(7)$~.

As one can see from the preceding discussion, all helicity amplitudes
or form factors (transverse, longitudinal and scalar) can be expressed
in terms of two types of elementary spatial matrix elements,
\ba
F(k) &=& \langle f|\hat U|i \rangle ~,
\nonumber\\
G_m(k) &=& \langle f|\hat T_m|i \rangle ~.
\ea
These matrix elements contain all the hadron structure information.
They are very sensitive to the hadron wave function. In photo- and
electroproduction of baryon resonances only form factors in which the
initial state is the nucleon (proton or neutron) can be measured.
In Appendix~D we show explicitly how $F(k)$ and $G_m(k)$ can be evaluated
in $U(7)$ ~.

\section{Form factors in photo- and electroproduction}
\setcounter{equation}{0}

The important form factors in photo- and electroproduction are those
connecting the ground state (the nucleon, proton or neutron) to its
excited states. These form factors can be evaluated in explicit form
in three different cases.

(i) Harmonic oscillator quark model.\\
In Table~\ref{ho} we present the elementary form factors,
$F(k)$ and $G_m(k)$, in the harmonic oscillator quark model. These form
factors had been obtained previously \cite{CKO,CL} in closed analytic
form and are reported here for comparison with those of the collective
model to be discussed next. The elastic form factor is given by
\ba
F(k) &=& \mbox{e}^{- k^2 \beta^2/6} ~.
\ea
The scale parameter $\beta$ is related to the harmonic oscillator
size parameter $\alpha$ by $\beta=1/\alpha$.

(ii) Collective model: end string.\\
We consider first the case in which the charge and magnetic moment is
concentrated at the end points of the string of Figure~\ref{geometry}.
For this case, the elastic form factor is given in terms of a
spherical Bessel function (see Appendix~D)
\ba
F(k) &=& j_0(k \beta) ~.
\ea

(iii) Collective model: distributed string.\\
We consider now the case in which the charge and magnetic moment are
distributed along the strings of Figure~\ref{geometry} with a probability
distribution
\ba
g(\beta) &=& \beta^2 \, \mbox{e}^{-\beta/a} ~. \label{prob}
\ea
In this case the form factors are obtained by integrating the form factors
of case (ii). For the elastic form factor one has
\ba
F(k) \;=\; \left. \int_0^{\infty} g(\beta) j_0(k \beta) \,
\mbox{d} \beta \right/ \int_0^{\infty} g(\beta) \mbox{d} \beta
\;=\; \frac{1}{(1+k^2a^2)^2} ~.
\ea
In Tables~\ref{string1} and~\ref{string2} we present analytic results for
several transition form factors of the collective model for the end string
and the distributed string, respectively (see Appendix~D).

Comparing the elastic form factors for the three cases
\ba
F(k) \;=\; \left\{ \begin{array}{lll}
\mbox{e}^{- k^2 \beta^2/6} & \mbox{harmonic oscillator} &
(R^2=0, \; N \rightarrow \infty) \\ & & \\
j_0(k \beta) & \mbox{end string} &
(R^2>0, \; N \rightarrow \infty) \\ & & \\
1/(1+k^2 a^2)^2 & \mbox{distributed string} &
(R^2>0, \; N \rightarrow \infty) \end{array} \right. \label{fk}
\ea
one can see that those of the string model drop as a power of $k$,
while those of the harmonic oscillator drop exponentially with $k$.
It is a major property of hadrons that form factors fall off as powers
of $k$. This power behavior is naturally obtained in a collective string
model, although the end string oscillates,
$j_0(k \beta)=\sin(k \beta)/k \beta$.
In order to obtain a similar behavior in quark models, a quark form
factor is introduced and/or the wave function is boosted \cite{LP,FH}.

We also note that all form factors are given in terms of one parameter
describing the size of the hadron. This parameter can be determined
from a measurement of the r.m.s. radius of the hadron.
For $k \rightarrow 0$ one has
\ba
F(k) \;\rightarrow\; \left\{ \begin{array}{l}
1-k^2 \beta^2/6 + \ldots \\ \\
1-k^2 \beta^2/6 + \ldots \\ \\
1-2 k^2 a^2 + \ldots \end{array} \right. \label{fk0}
\ea
which gives the relation between the parameters, $\beta$ and
$a$, and the r.m.s. radius.

\section{Comparison with experimental helicity amplitudes}
\setcounter{equation}{0}

The final step in comparing with experimental data is the choice of
reference frame, which determines the relation between
$k^2$ and $Q^2=k^2-k_0^2$. We perform our calculations in
the equal momentum or Breit frame (in which recoil contributions vanish),
\ba
k^2 &=& Q^2 + \frac{(W^2-M^2)^2}{2(M^2+W^2)+Q^2} ~.
\ea
$M$ is the nucleon mass, $W$ is the mass of the resonance and
$-Q^2=k_0^2-k^2$ can be interpreted as the mass squared of the virtual
photon.

In photoproduction one has $Q^2=0$ and hence the photon momentum $k=k_0$.
With this value of the momentum transfer $k$ and the formalism of the
preceding sections, we calculate the transverse helicity amplitudes for
photoproduction. The results are presented in Tables~\ref{nphoto}
and~\ref{dphoto}. Calculation (1) is the harmonic oscillator result,
whereas calculations (2) and (3) are the results of the distributed
string with $R^2=0.5$ and $R^2=1.0$, respectively. (We do not quote
the results of the end string.) There are no free
parameters in the calculation with the exception of a size parameter,
$\beta$ for (1) and $a$ for (2) and (3). This size parameter is fixed
by the value of the proton r.m.s. radius,
$\langle r^2 \rangle^{1/2} =0.862 \pm 0.012$ fm \cite{prms}.
The last column in Tables~\ref{nphoto} and~\ref{dphoto} shows the
experimental helicity amplitudes, quoted in \cite{PDG} with a sign.
This sign can neither be extracted from the data independently from the
sign of the subsequent decay amplitude of the resonance
($\mbox{N}^* \rightarrow \mbox{N} + \pi$)
nor can be determined by calculations. It is thus a conventional sign
(except for the relative sign of amplitudes leading to the same final state).
In Tables~\ref{nphoto} and~\ref{dphoto} we have used the signs of the
harmonic oscillator limit (with the conventions of \cite{MM} for the
harmonic oscillator wave functions), since the purpose of the present paper
is to investigate different scenarios of hadronic structure in as much
as possible model independent way. We note, however, that the sign
conventions used by previous authors are often in disagreement with one
another. In particular, the sign conventions of \cite{Capstick} are in
disagreement with those of \cite{KI} and \cite{CL}. For example, in
\cite{CL} the nonrelativistic contribution to the amplitude $A^p_{1/2}$
leading to the state N$(1720)P_{13}$ is calculated to be $-113$, while
in \cite{Capstick} it is calculated to be $+112$. The comparison between
theory and experiment in Tables~\ref{nphoto} and~\ref{dphoto} should
therefore be restricted to absolute values of the amplitudes (and their
relative phases when leading to the same final states).
With this in mind, the agreement
with experimental data is fair and it is approximately the
same for all calculations. The reason is that the photocouplings
depend almost completely on the spin-flavor part. The dependence
on the spatial part is minor, since the helicity amplitudes are
evaluated at a relatively small momentum $k=k_0$, for which all
calculations given similar results. To emphasize this point, consider
the helicity amplitudes for the excitation of the $\Delta(1232)P_{33}$
resonance. These can be written as
\ba
A_{1/2} &=& -\frac{2}{3} \sqrt{\frac{2\pi}{k_0}} \mu k \, F(k) ~,
\nonumber\\
A_{3/2} &=& -\frac{2}{3} \sqrt{\frac{6\pi}{k_0}} \mu k \, F(k) ~,
\ea
where $F(k)$ is given by Eq.~(\ref{fk}). One first observes that the
ratio $A_{3/2}/A_{1/2} = \sqrt{3}$ is independent of $k$. This result is due
to spin-flavor symmetry and is in good agreement with the experimental
value $1.83 \pm 0.15$. Moreover, for small $k$ the form factors
are essentially identical, as one can see directly from Eq.~(\ref{fk0}).

We stress the fact that in the calculations we have insisted on
describing correctly the proton r.m.s. radius. If this condition
is relaxed, and $\beta$ (or $a$) is used as a parameter, a better
description of the photocouplings can be obtained.
The description can be improved further by changing the values of the
$g$-factor (and accordingly the quark effective mass), as done in
\cite{Capstick}.

There are several places where there is a large disagreement between
experiment and calculation. We mention here in particular the pair of
states N$(1535)S_{11}$ and N$(1650)S_{11}$. The disagreement in this case
can be corrected by mixing the two states \cite{HLC,FH}
\ba
|\mbox{N}(1535)S_{11} \rangle &=&+|^{2}8_{1/2} \rangle \, \cos(-38^{\circ})
+ |^{4}8_{1/2} \rangle \, \sin(-38^{\circ}) ~,
\nonumber\\
|\mbox{N}(1650)S_{11} \rangle &=&-|^{2}8_{1/2} \rangle \, \sin(-38^{\circ})
+ |^{4}8_{1/2} \rangle \, \cos(-38^{\circ}) ~. \label{mix}
\ea
This mixing breaks $SU_{sf}(6)$ and can only be introduced by a
tensor-like interaction, Eq.~(\ref{tensor}). It appears to be the only
place in the spectrum and transitions where there is a clear evidence
of such a breaking.

By comparing the photocouplings of the
N$(1440)P_{11}$ and the N$(1710)P_{11}$ resonances for the
harmonic oscillator and the collective string we see that in this case
there is a clear difference between the two models.
This is due to the different nature of these states: in the harmonic
oscillator these states belong to the two-phonon multiplet whereas
in the collective string they correspond to the fundamental (one-phonon)
vibrations of the string.

Another observation is that the N$(1520)D_{13}$ and N$(1680)F_{15}$
resonances are predominantly excited with helicity $\mu=3/2$ for proton
targets. The $A^p_{1/2}$ amplitude is small because of a cancellation of
the magnetic and electric contributions. Tables~\ref{ho}--\ref{string2}
show that this feature is present in both the
collective string and the harmonic oscillator model.
It appears to be a consequence of the spin-flavor symmetry and
does not depend on details of the hadron structure.

Next we turn to a discussion of the transition form factors as are
measured in electroproduction. It is here that major differences
between the various models of hadron structure occur, since the
form factors discussed in Section~9 differ widely for large $k^2$
(or $Q^2$). In Figure~\ref{protonff} we show a comparison between the
experimental data for the proton elastic form factor and the form factors
of Eq.~(\ref{fk}). One can see clearly that only the form factor for the
distributed string (dipole form factor) describes the data well.
Similarly, we show in Figure~\ref{deltaff} the transition form factor
exciting the $\Delta(1232)P_{33}$ resonance. Again it appears that the
distributed string provides the best description.
Figure~\ref{deltaff} also indicates there is an additional contribution
to $G_M^{\Delta}$ of magnitude $\sim 0.2 \times 3F_D$, which falls off
with $Q^2$ faster than the leading order $(Q^{-4}$).

Figures~\ref{d13}--\ref{f15} show the helicity amplitudes for the
N$(1520)D_{13}$, N$(1535)S_{11}$, N$(1650)S_{11}$ and N$(1680)F_{15}$
resonances. Although the experimental information is not very accurate,
again it appears that the distributed string form factors decribe the
observed data better.

Since on one side the accurate measurement of transition form factors is
part of the experimental programs at ELSA, MAMI and CEBAF
and on the other side it is possible, with the methods
discussed in this article, to compute any transition form factor,
we believe that when the measurements will be completed, we will be
able to make more definitive statements concerning the structure of the
nucleon and its resonances and the validity of the various models
of hadronic structure.

Finally, we note that, while the individual helicity amplitudes test
models of hadronic structure, the helicity asymmetries
\ba
\frac{A_{1/2}^2-A_{3/2}^2}{A_{1/2}^2+A_{3/2}^2}
\ea
test more the form of the transition operator and the spin-flavor part
of the wave functions. This is shown in
Figures~\ref{asymd13} and~\ref{asymf15} where it is seen that both the
harmonic oscillator quark model and the distributed string model give
essentially the same results for the helicity asymmetries of the
N$(1520)D_{13}$ and the N$(1680)F_{15}$ resonances.

\section{Conclusions}
\setcounter{equation}{0}

We have presented here a framework within which different scenarios
for baryon structure in the nonperturbative regime can be studied.
This framework makes use of a spectrum generating algebra
\ba
{\cal  G} &=& U(7) \otimes SU_{sf}(6) \otimes SU_{c}(3) ~.
\ea
In particular, we have analyzed two different cases: (i) a
`single-particle' harmonic oscillator model and (ii) a `collective'
string model. We find that most results are independent of which model
is used for the spatial part and are a consequence of the spin-flavor
structure of the problem,
$SU_{sf}(6) \supset SU_{f}(3) \otimes SU_{s}(2)$,
and of the triality property of baryons, namely of the fact that there
are three constituent parts which transform as the triplet representation
of $SU_{c}(3)$. Differences between models occur only in the
transition form factors which provide therefore an important clue to
understand baryon structure. Present data suggest that the collective
model with charge, mass and magnetic moment distributed along the
string is the most likely description. New and more accurate data
which can be obtained at new facilities like ELSA, MAMI and CEBAF
may help elucidating the situation.
The fact that the formalism has been set up in a as much as possible
model independent way also gives the possibility to search for `new'
physics, which in this context means unconventional configurations of
quarks and gluons.

The analysis presented in this article has been carried out assuming
$S_3$ (or $D_3$) symmetry ({\it i.e.} three identical constituents with
identical interactions). The next step in this study is the breaking of
this symmetry. This is due to two effects: (i) the mass of the three
constituents may not be the same; this breaking will allow us to
treat strange baryons as well, and (ii) the interaction between the
three objects may be such that the geometric arrangement is not that
of an equilateral triangle with $D_3$ symmetry; this breaking will
allow us to discuss the electric form factor of the neutron, which
is identically zero if the neutron has $S_3$ symmetry. Both these
types of symmetry breaking can be studied in the framework of the
present formalism and will be discussed in more detail in the next
publication in this series.

\section*{Acknowledgements}

This work is supported in part (RB) by the Stichting voor Fundamenteel
Onderzoek der Materie (FOM) with financial support from the Nederlandse
Organisatie voor Wetenschappelijk Onderzoek (NWO), (FI) by D.O.E. grant
DE-FG02-91ER40608 and (AL) by the Israeli Science Ministry and by the
Basic Research Foundation of the Israel Academy of Sciences and Humanities.

\appendix
\section{Basic matrix elements}
\setcounter{equation}{0}

The matrix elements of the mass-squared operator of Eq.~(\ref{ms3})
(and of all other $U(7)$ operators of interest) can be calculated in
either one of the two sets of basis states discussed in Section~4.
With standard Racah algebra \cite{Talmi} they can be expressed in terms
of the reduced matrix elements of the boson creation and annihilation
operators themselves.\\
(i) In the harmonic oscillator $U(4) \supset U(3)$ basis we have
\ba
\langle N,n^{\prime},L^{\prime}||b^{\dagger}||N-1,n,L \rangle &=&
\langle N-1,n,L||\tilde{b}||N,n^{\prime},L^{\prime} \rangle
\nonumber\\
&=& \delta_{n^{\prime},n+1} \, \left\{ \begin{array} {ll}
\sqrt{(n+L+3)(L+1)} & \mbox{ for } L^{\prime}=L+1 ~, \\ & \\
\sqrt{(n-L+2)L}     & \mbox{ for } L^{\prime}=L-1 ~, \end{array} \right.
\nonumber\\
\langle N,n^{\prime},L^{\prime}|| s^{\dagger}||N-1,n,L \rangle &=&
\langle N-1,n,L||\tilde{s}||N,n^{\prime},L^{\prime} \rangle
\nonumber\\
&=& \delta_{n^{\prime},n} \delta_{L^{\prime},L} \, \sqrt{N-n} ~.
\ea
(ii) In the $U(4) \supset SO(4)$ basis we have \cite{HP}
\ba
\langle N,\omega^{\prime},L^{\prime}||c^{\dagger}_l||N-1,\omega,L \rangle
&=& \langle N-1,\omega,L||\tilde{c}_l||N,\omega^{\prime},L^{\prime}
\rangle
\nonumber\\
&=& \sqrt{(2L^{\prime}+1)(2l+1)(2L+1)} \, \left\{ \begin{array}{ccc}
\omega^{\prime}/2 & \omega^{\prime}/2 & L^{\prime} \\
\omega/2 & \omega/2 & L \\ 1/2 & 1/2 & l \end{array} \right\}
\nonumber\\
&& \times
\langle N,\omega^{\prime} ||| c^{\dagger} ||| N-1,\omega \rangle ~,
\ea
where the $SO(4)$ reduced (triple bar) matrix elements are
\ba
\langle N,\omega^{\prime} ||| c^{\dagger} ||| N-1,\omega \rangle &=&
\left\{ \begin{array}{ll} \sqrt{\omega(\omega+1)(N-\omega+1)/2}
& \mbox{ for } \omega^{\prime}=\omega-1 ~, \\ & \\
\sqrt{(\omega+1)(\omega+2)(N+\omega+3)/2}
& \mbox{ for } \omega^{\prime}=\omega+1 ~. \end{array} \right.
\ea
Here $c_l^{\dagger}$ denotes for $l=0$ the scalar boson and
for $l=1$ the vector boson.

\section{Large N limit and coherent states}
\setcounter{equation}{0}

In the study of the properties of the string-like configuration
of Figure~\ref{geometry}, it is often convenient to introduce a
coherent (or intrinsic) state basis. The use of this basis
helps elucidating the physical content of algebraic models by
stressing their geometry, and allows one to obtain closed analytic
expressions for observables in the limit of large $N$. In view of
confinement we expect $N$ to be large, since $N$ determines
the number of bound states in the model.

Coherent states of $U(k)$ are described in \cite{cs}. Here we need
the coherent states of $U(7)$ (the coset space $U(7)/U(6) \otimes U(1)$).
They are characterized by six complex variables.
Removing overall rotations and considering static problems, one is left
with only three real variables, $(r_{\rho},r_{\lambda},\theta)$: two
radial coordinates and the angle inbetween.
These variables characterize the shape of an object composed of three
constituent parts and their geometric meaning is illustrated in
Figure~\ref{variables}. The general properties of $U(7)$
can be studied by using mean-field techniques.
For a system of boson degrees of freedom, the variational wave function is
a coherent state which take the form of a condensate of $N$ bosons,
which depends parametrically on the shape variables.
The expectation value of an algebraic mass operator in this condensate
defines a classical potential function $V(r_{\rho},r_{\lambda},\theta)$.
The equilibrium shape parameters are defined
by the global minimum of the potential function and are found by
a variational calculation. Substituting these equilibrium values in the
variational wave function provides an intrinsic state
representing the equilibrium shape.
For the $S_3$-invariant mass-squared operator of Eq.~(\ref{ms3}) the
equilibrium shape parameters
satisfy $r_{\rho}=r_{\lambda}>0$ and $\theta=\pi/2$ \cite{BL}.
These are precisely the conditions satisfied by the Jacobi coordinates of
Eq.~(\ref{jacobi}) for an equilateral triangular shape with a threefold
symmetry axis (in our convention along the $y$-axis).
In this case, the coherent state basis depends only on one variable,
$R=\sqrt{r^2_{\rho}+r^2_{\lambda}}$. It is obtained by introducing
the following boson operators
\ba
b_c^{\dagger} &=& \frac{1}{\sqrt{1+R^2}} \Bigl[ s^{\dagger}
+ R \frac{1}{\sqrt{2}} \Bigl( b_{\rho,z }^{\dagger}
+ b_{\lambda,x}^{\dagger} \Bigr) \Bigr] ~,
\nonumber\\
b^{\dagger}_{u} &=& \frac{1}{\sqrt{1+R^2}} \Bigl[ - R \, s^{\dagger}
+ \frac{1}{\sqrt{2}} \Bigl( b_{\rho,z }^{\dagger}
+ b_{\lambda,x}^{\dagger} \Bigr) \Bigr] ~,
\nonumber\\
b^{\dagger}_{v} &=& \frac{1}{\sqrt{2}}
\Bigl( b_{\rho,z}^{\dagger} - b_{\lambda,x}^{\dagger} \Bigr) ~,
\nonumber\\
b^{\dagger}_{w} &=& \frac{1}{\sqrt{2}}
\Bigl( b_{\rho,x}^{\dagger} + b^{\dagger}_{\lambda,z} \Bigr) ~,
\nonumber\\
b^{\dagger}_{1} &=& \frac{1}{\sqrt{2}}
\Bigl( b^{\dagger}_{\rho,y} + b^{\dagger}_{\lambda,y} \Bigr) ~,
\nonumber\\
b^{\dagger}_{2} &=& \frac{1}{\sqrt{2}}
\Bigl( - b^{\dagger}_{\rho,x} + b_{\lambda,z}^{\dagger} \Bigr) ~,
\nonumber\\
b^{\dagger}_{3} &=& \frac{1}{\sqrt{2}}
\Bigl( b^{\dagger}_{\rho,y} -  b^{\dagger}_{\lambda,y} \Bigr) ~.
\label{bvib}
\ea
The ground state of a system with $S_3$ symmetry is represented by
a condensate of the form
\ba
| N;R \rangle_c &=& \frac{1}{\sqrt{N!}}
\Bigl( b^{\dagger}_{c} \Bigr)^{N} |0 \rangle ~.
\label{bc}
\ea
This state contains several angular momentum states, since for $R>0$ the
boson operator $b^{\dagger}_c$ is a mixture of scalar and vector bosons.
States with good quantum numbers are obtained by projection.
For $R=0$ the boson condensate of
Eq.~(\ref{bc}) becomes the ground state of the harmonic oscillator
\ba
| N;R=0 \rangle &=& \frac{1}{\sqrt{N!}}
\Bigl( s^{\dagger} \Bigr)^{N} |0 \rangle ~,
\ea
which contains only a single state with $L^{\pi}=0^+$.

Excited states are obtained by replacing condensate bosons by other members
of the basis Eq.~(\ref{bvib}).
Vibrational excitations are represented by the $b^{\dagger}_u$,
$b^{\dagger}_v$, $b^{\dagger}_w$ boson operators \cite{BL}, and are
written as
\ba
\frac{1}{\cal N} \,
\Bigl( b^{\dagger}_{u} \Bigr)^{n_u}
\Bigl( b^{\dagger}_{v} \Bigr)^{n_v}
\Bigl( b^{\dagger}_{w} \Bigr)^{n_w}
\Bigl( b^{\dagger}_{c} \Bigr)^{N-n_u-n_v-n_w} |0 \rangle ~, \label{cuvw}
\ea
where ${\cal N}$ is a normalization constant.
The three rotational bosons $b^{\dagger}_1$,
$b^{\dagger}_2$, $b^{\dagger}_3$ represent instead spurious excitations
of the condensate, corresponding to overall rotations (Goldstone modes).

\section{Spin-flavor wave functions}
\setcounter{equation}{0}

In this appendix we list the conventions used for the spin and
flavor wave functions.\\
(i) Spin wave functions \cite{KI}:
\ba
\begin{array}{ccl}
S=1/2 && |\chi^{\rho}_{1/2} \rangle \;=\;
( |\uparrow \downarrow \uparrow \rangle
- |\downarrow \uparrow \uparrow \rangle)/\sqrt{2} ~, \\ && \\
&& |\chi^{\lambda}_{1/2} \rangle \;=\;
(2|\uparrow \uparrow \downarrow \rangle
- |\uparrow \downarrow \uparrow \rangle
- |\downarrow \uparrow \uparrow \rangle)/\sqrt{6} ~, \\ && \\
S=3/2 && |\chi^S_{3/2} \rangle \;=\; |\uparrow \uparrow \uparrow \rangle ~.
\end{array}
\ea
We only show the state with the largest component of the projection
$M_S=S$. The other states are obtained by applying the lowering operator
in spin space.\\
(ii) Flavor wave functions \cite{KI}:
\ba
\begin{array}{ccl}
\mbox{octet} && |\phi_p^{\rho} \rangle \;=\;
(|udu \rangle-|duu \rangle)/\sqrt{2} ~, \\ && \\
&& |\phi_p^{\lambda} \rangle \;=\;
(2|uud \rangle-|udu \rangle-|duu \rangle)/\sqrt{6} ~, \\ && \\
\mbox{decuplet} && |\phi^S_{\Delta^{++}} \rangle \;=\;
|uuu \rangle ~. \end{array}
\ea
We only show the highest charge state. The other charge states
are obtained by applying the lowering operator in isospin space.

\section{Radial integrals}
\setcounter{equation}{0}

All helicity amplitudes in $U(7)$ are expressed in terms
of two types of elementary spatial matrix elements (or radial integrals)
\ba
F(\epsilon) &=& \langle f|\hat U|i \rangle \;=\;
\langle f|\mbox{e}^{ -i \epsilon \hat D_{\lambda,z} }|i \rangle ~,
\nonumber\\
G_m(\epsilon) &=& \langle f|\hat T_m|i \rangle
\;=\; \frac{i}{2} \eta \, \langle f|
\hat D_{\lambda,m} \, \mbox{e}^{ -i \epsilon \hat D_{\lambda,z} }
+ \mbox{e}^{ -i \epsilon \hat D_{\lambda,z} } \, \hat D_{\lambda,m}
|i \rangle ~.
\label{fg}
\ea
Here $\epsilon=k \beta/X_D$ and $\eta=m_q k_0 \beta/X_D$
(see Eq.~(\ref{emop})).
These matrix elements can be calculated exactly in $U(7)$
by using the symmetry properties of the transition operator.
We first discuss how they are calculated in general. Next we
discuss two special cases for which they are derived in
closed analytic form.

\subsection{General case}

The most convenient basis to derive a general expression for the
spatial matrix elements is the second one discussed in Section~4,
\ba
|N,(n_\rho,L_{\rho}),(\omega,L_{\lambda});L,M_L \rangle ~.
\ea
Since the operator $\hat D_{\lambda,z}$ appearing in the exponent
is a generator of $SO_{\lambda}(4)$, its matrix elements are
diagonal in $\omega$. Moreover, since $\hat D_{\lambda,m}$  does not
depend on the $\rho$--coordinate, its matrix elements are also diagonal
in the harmonic oscillator labels ($n_{\rho},L_{\rho}$).
The matrix elements of $\hat U$ can be evaluated by decoupling the
$\rho$--part
\ba
F(\epsilon) &=& \langle N,(n_{\rho}^{\prime},L_{\rho}^{\prime}),
(\omega^{\prime},L_{\lambda}^{\prime});L^{\prime},M_L^{\prime}|
\mbox{e}^{ -i \epsilon \hat D_{\lambda,z} }
|N,(n_{\rho},L_{\rho}),(\omega,L_{\lambda});L,M_L \rangle
\nonumber\\
&=& \delta_{n_{\rho}^{\prime},n_{\rho}}
\delta_{L_{\rho}^{\prime},L_{\rho}}
\sum_{M_{\rho},M_{\lambda},M_{\lambda}^{\prime}}
\langle L_{\rho},M_{\rho},L_{\lambda}^{\prime},M_{\lambda}^{\prime}
|L^{\prime},M_L^{\prime} \rangle
\langle L_{\rho},M_{\rho},L_{\lambda},M_{\lambda}|L,M_L \rangle
\nonumber\\
&& \times \langle N-n_{\rho},\omega^{\prime},L_{\lambda}^{\prime},
M_{\lambda}^{\prime}|\mbox{e}^{ -i \epsilon \hat D_{\lambda,z} }
|N-n_{\rho},\omega,L_{\lambda},M_{\lambda} \rangle ~,
\ea
and by recognizing that the matrix elements appearing in the r.h.s.
can be interpreted as representation matrix elements of $SO(4)$.
(The projection of the angular momentum $M_{\rho}$, $M_{\lambda}$
should not be confused with $M_{\rho}$, $M_{\lambda}$ representation
of $S_3$.) They can be derived by using the isomorphism between
$SO(4)$ and $SU(2) \otimes SU(2)$ \cite{BA2}
\ba
\langle N,\omega^{\prime},L^{\prime},M^{\prime}_L|
\mbox{e}^{ -i \epsilon \hat D_{\lambda,z} }
|N,\omega,L,M_L \rangle \;=\; \hspace{6cm}
\nonumber\\
\delta_{\omega^{\prime},\omega} \delta_{M_L^{\prime},M_L} \sum_{\mu}
\left< \frac{\omega}{2},\mu,\frac{\omega}{2},M_L-\mu \Big|
L,M_L \right>
\left< \frac{\omega}{2},\mu,\frac{\omega}{2},M_L-\mu \Big|
L^{\prime},M_L \right>
\nonumber\\
\times \left[
\frac{1+(-1)^{L+L^{\prime}}}{2} \cos [(2\mu-M_L)\epsilon] - i \,
\frac{1-(-1)^{L+L^{\prime}}}{2} \sin [(2\mu-M_L)\epsilon] \right] ~.
\label{ume}
\ea
The matrix elements of $\hat T_{m}$ are obtained in a similar way.
Again we decouple the $\rho$-part of the wave function and
then we insert a complete set of intermediate states to obtain
\ba
G_m(\epsilon) &=& \langle N,(n_{\rho}^{\prime},L_{\rho}^{\prime}),
(\omega^{\prime},L_{\lambda}^{\prime});L^{\prime},M_L^{\prime}|
\hat T_m |N,(n_{\rho},L_{\rho}),(\omega,L_{\lambda});L,M_L \rangle
\nonumber\\
&=& \frac{i}{2} \eta \, \delta_{n_{\rho}^{\prime},n_{\rho}}
\delta_{L_{\rho}^{\prime},L_{\rho}}
\sum_{M_{\rho},M_{\lambda},M_{\lambda}^{\prime}}
\langle L_{\rho},M_{\rho},L_{\lambda}^{\prime},M_{\lambda}^{\prime}
|L^{\prime},M_L^{\prime} \rangle
\langle L_{\rho},M_{\rho},L_{\lambda},M_{\lambda}|L,M_L \rangle
\nonumber\\
&& \times \sum_{L_{\lambda}^{\prime\prime}} \left[ \langle
N-n_{\rho},\omega^{\prime},L_{\lambda}^{\prime},M_{\lambda}^{\prime}|
\hat D_{\lambda,m} |N-n_{\rho},\omega,
L_{\lambda}^{\prime\prime},M_{\lambda} \rangle \right.
\nonumber\\
&& \hspace{2cm} \times
\langle N-n_{\rho},\omega,L_{\lambda}^{\prime\prime},M_{\lambda}|
\mbox{e}^{ -i \epsilon \hat D_{\lambda,z} }
|N-n_{\rho},\omega,L_{\lambda},M_{\lambda} \rangle
\nonumber\\
&& \hspace{1cm} + \langle
N-n_{\rho},\omega^{\prime},L_{\lambda}^{\prime},M_{\lambda}^{\prime}|
\mbox{e}^{ -i \epsilon \hat D_{\lambda,z} }
|N-n_{\rho},\omega^{\prime},L_{\lambda}^{\prime\prime},M_{\lambda}^{\prime}
\rangle
\nonumber\\
&& \hspace{2cm} \times \left.
\langle N-n_{\rho},\omega^{\prime},L_{\lambda}^{\prime\prime},
M_{\lambda}^{\prime}| \hat D_{\lambda,m}
|N-n_{\rho},\omega,L_{\lambda},M_{\lambda} \rangle
\right] ~.
\ea
The matrix elements of $\hat U$ are given in Eq.~(\ref{ume}), whereas
those of $\hat D_{\lambda,m}$ can be expressed in terms of their
reduced matrix elements by using the Wigner-Eckart theorem
\ba
\langle N,\omega^{\prime},L^{\prime},M^{\prime}| \hat D_{m}
|N,\omega,L,M \rangle
&=& \frac{\langle L,M,1,m|L^{\prime},M^{\prime} \rangle}
{\sqrt{2L^{\prime}+1}} \,
\langle N,\omega^{\prime},L^{\prime}|| \hat D || N,\omega,L \rangle ~,
\ea
where
\ba
\langle N,\omega^{\prime},L^{\prime} || \hat D || N,\omega,L \rangle &=&
\langle N,\omega,L || \hat D || N,\omega^{\prime},L^{\prime} \rangle
\nonumber\\
&=& \delta_{\omega^{\prime},\omega} \sqrt{(\omega-L)(\omega+L+2)(L+1)}
\hspace{1cm} \mbox{for } L^{\prime}=L+1 ~.
\ea
The matrix element of $\hat T_z$ can also be
obtained in a more direct way by noting that
\ba
G_0(\epsilon) &=& - \eta \frac{\mbox{d}F(\epsilon)}{\mbox{d}\epsilon} ~.
\ea

\subsection{Harmonic oscillator}

There exist special cases in which the
matrix elements of the electromagnetic transition operator
can be derived in closed analytic form. In this subsection we show
how the well-known results of the harmonic oscillator quark model
are obtained in $U(7)$~.
In this case, the most appropriate basis is the first one discussed
in Section~4, namely
\ba
|N,(n_{\rho},L_{\rho}),(n_{\lambda},L_{\lambda});L,M_L \rangle ~.
\ea
In the harmonic oscillator limit, the spatial part of the ground state
wave function is
\ba
|[56,0^+]_0,0 \rangle &\equiv& |N,(0,0),(0,0),0,0 \rangle ~.
\ea
The subscript indicates the harmonic oscillator shell quantum number
$n_{\rho}+n_{\lambda}$.
The spatial wave functions for some of the lowest excited states are
\ba
|[70,1^-_{\lambda}]_1,M_L \rangle &\equiv& |N,(0,0),(1,1),1,M_L \rangle ~,
\nonumber\\
|[56,0^+]_2,0 \rangle &\equiv& \frac{1}{\sqrt{2}}
(|N,(2,0),(0,0),0,0 \rangle+|N,(0,0),(2,0),0,0 \rangle) ~,
\nonumber\\
|[70,0^+_{\lambda}]_2,0 \rangle &\equiv& \frac{1}{\sqrt{2}}
(|N,(2,0),(0,0),0,0 \rangle-|N,(0,0),(2,0),0,0 \rangle) ~,
\nonumber\\
|[56,2^+]_2,M_L \rangle &\equiv& \frac{1}{\sqrt{2}}
(|N,(2,2),(0,0),2,M_L \rangle+|N,(0,0),(2,2),2,M_L \rangle) ~,
\nonumber\\
|[70,2^+_{\lambda}]_2,M_L \rangle &\equiv& \frac{1}{\sqrt{2}}
(|N,(2,2),(0,0),2,M_L \rangle-|N,(0,0),(2,2),2,M_L \rangle) ~.
\ea
Just as in the general case, the
matrix elements of $\hat U$ and $\hat T_m$ for the excitation
of a baryon resonance from the ground state (nucleon) can
be expressed in terms of $U(4)$ matrix elements by decoupling
the $\rho$--part
\ba
\langle N,(n_{\rho},L_{\rho}),(n_{\lambda},L_{\lambda});L,M_L|
\left( \begin{array}{l} \hat U \\ \hat T_m \end{array}
\right) |N,(0,0),(0,0),0,0 \rangle \;= \hspace{2cm}
\nonumber\\
\delta_{n_{\rho},0} \delta_{L_{\rho},0}
\langle N,n_{\lambda},L_{\lambda},M_L| \left( \begin{array}{l} \hat U \\
\hat T_m \end{array} \right) |N,0,0,0 \rangle ~.
\ea
The r.h.s. represents matrix elements in the harmonic oscillator
$U(4) \supset U(3)$ basis which are given by \cite{BAS}
\ba
\langle N,n,L,M_L| \hat U |N,0,0,0 \rangle &=& \delta_{M_L,0}
(-)^{(n-L)/2} \sqrt{\frac{N!(2L+1)}{(N-n)!(n+L+1)!!(n-L)!!}}
\nonumber\\
&& \times (\cos \epsilon)^{N-n} (-i \, \sin \epsilon)^n ~,
\nonumber\\
\langle N,n,L,M_L| \hat T_z |N,0,0,0 \rangle &=&
i \, \eta \, \delta_{M_L,0}
(-)^{(n-L)/2} \sqrt{\frac{N!(2L+1)}{(N-n)!(n+L+1)!!(n-L)!!}}
\nonumber\\
&& \times (\cos \epsilon)^{N-n-1} (-i \, \sin \epsilon)^{n-1}
(n-N \sin^2 \epsilon) ~,
\nonumber\\
\langle N,n,L,M_L| \hat T_+ |N,0,0,0 \rangle &=&
-i \, \eta \, \delta_{M_L,1}
(-)^{(n-L)/2} \sqrt{\frac{N!L(L+1)(2L+1)}{(N-n)!(n+L+1)!!(n-L)!!}}
\nonumber\\
&& \times (-i \, \sin \epsilon)^{n-1} \, \frac{1}{2}
\left[ (\cos \epsilon)^{N-n+1} + (\cos \epsilon)^{N-n} \right] ~.
\ea
We have used the sign conventions of \cite{MM} for the harmonic
oscillator wave functions.

The elastic form factor is then given by
\ba
F(\epsilon) \;=\; \langle [56,0^+]_0,0| \hat U |[56,0^+]_0,0 \rangle
\;=\; (\cos \epsilon)^N \;\rightarrow\; \mbox{e}^{-k^2 \beta^2/6} ~.
\ea
In the last step we have taken the large $N$ limit, which is such that
$n \ll N$ (where $n$ denotes the oscillator shell), and
$\epsilon \sqrt{3N}(=k \beta)$ remains finite. For the harmonic
oscillator we have $\epsilon=k \beta/X_D=k \beta/\sqrt{3N}$.
For the inelastic transition to the first excited negative parity
state we find
\ba
G_+(\epsilon) \;=\;
\langle [70,1^-_{\lambda}]_1,1| \hat T_+ |[56,0^+]_0,0 \rangle
&=& -i \, \eta \sqrt{N/2} \,
\left[ (\cos \epsilon)^{N} + (\cos \epsilon)^{N-1} \right]
\nonumber\\
&\rightarrow& -i \, \sqrt{\frac{2}{3}} \, m_q k_0 \beta \,
\mbox{e}^{-k^2 \beta^2/6} ~.
\ea
We have used $\eta=m_q k_0 \beta/X_D=m_q k_0 \beta/\sqrt{3N}$.
{}From a comparison with the expressions derived in coordinate space
we find that the scale parameter $\beta$ is inversely proportional to
the harmonic oscillator size parameter $\alpha=1/\beta$.

In conclusion, we find that for $N \rightarrow \infty$ the form
factors in the harmonic oscillator limit of $U(7)$ are
identical to those in the harmonic oscillator quark model.
The results are summarized in Table~\ref{ho}.

\subsection{Collective model}

Also for the collective model discussed in Section~6 we can
derive closed analytic expressions for some of the form factors.
All rotational states belonging to the ground state band which has
$(n_u,n_v+n_w)=(0,0)$ can be obtained from the ground state
condensate of Eq.~(\ref{bc})
\ba
|N;R \rangle_c &=& \frac{1}{\sqrt{N!}}
\Bigl( b^{\dagger}_{c} \Bigr)^{N} |0 \rangle ~,
\ea
by projecting onto states of good angular momentum $L$
and good permutation symmetry.
Since the $L^{\pi}=0^+$ ground state is the only state with
$L=0$, one only has to project onto good angular momentum
\ba
| [56,0^+]_{(0,0)},0 \rangle &\equiv&
\int \mbox{d}\Omega \, |N;R,\Omega \rangle_c \; Y_{00}(\Omega) ~.
\ea
The subscript here denotes the vibrational quantum numbers
$(n_u,n_v+n_w)$ and $Y_{LM_L}(\Omega)$ are the usual spherical harmonics.

The elastic form factor is then given by
\ba
F(\epsilon) &=&
\langle [56,0^+]_{(0,0)},0| \hat U |[56,0^+]_{(0,0)},0 \rangle
\nonumber\\
&=& \int \mbox{d}\Omega \mbox{d}\Omega^{\prime}
\, Y^{\ast}_{00}(\Omega^{\prime}) \; _c\langle N;R,\Omega^{\prime} |
e^{-i \, \epsilon \hat D_{\lambda,z}} |N;R,\Omega \rangle_c \;
Y_{00}(\Omega) ~.
\ea
Here the angle $\Omega$ denotes the orientation of the condensate.
For $N \rightarrow \infty$ the matrix element in the integrand
is diagonal in $\Omega$ \cite{BA1} and we find
\ba
F(\epsilon) &\rightarrow& \frac{1}{4\pi} \int \mbox{d}\Omega
\; _c\langle N;R,\Omega | e^{-i \, \epsilon \hat D_{\lambda,z}}
|N;R,\Omega \rangle_c
\nonumber\\
&\rightarrow& \frac{1}{2} \int \mbox{d}(\cos \theta) \,
e^{-i \, \epsilon [N R \sqrt{2}/(1+R^2)] \cos \theta}
\nonumber\\
&=& j_0(k \beta) ~.
\ea
We have used that $\epsilon=k \beta/X_D$ with $X_D=N R \sqrt{2}/(1+R^2)$
for $N \rightarrow \infty$.
In a similar way we find for the transition to the first excited
negative parity state
\ba
G_+(\epsilon) &=& \langle [70,1^-_{\lambda}]_{(0,0)},1|
\hat T_+ |[56,0^+]_{(0,0)},0 \rangle
\nonumber\\
&=& \int \mbox{d}\Omega \mbox{d}\Omega^{\prime} \,
Y^{\ast}_{11}(\Omega^{\prime}) \; _c\langle N;R,\Omega^{\prime} |
\hat T_+ |N;R,\Omega \rangle_c \; Y_{00}(\Omega)
\nonumber\\
&\rightarrow& -i \, \eta \sqrt{2} \,
\frac{N R \sqrt{2}}{1+R^2} \, \frac{1}{4} \sqrt{3} \,
\int \mbox{d}(\cos \theta) \, (1-\cos^2 \theta) \,
e^{-i \, \epsilon [N R \sqrt{2}/(1+R^2)] \cos \theta}
\nonumber\\
&=& -i \, \sqrt{\frac{2}{3}} \, m_q k_0 \beta \,
[j_0(k \beta)+ j_2(k \beta)] ~.
\ea
For large $N$ the intrinsic state for the first excited vibrational band
with $(n_u,n_v+n_w)=(1,0)$, the bandhead of which we have
associated with the N$(1440)P_{11}$ Roper and the $\Delta(1600)P_{33}$
resonances, is given by
\ba
|N;R \rangle_u &=& b^{\dagger}_{u} \frac{1}{\sqrt{(N-1)!}}
\Bigl( b^{\dagger}_{c} \Bigr)^{N-1} |0 \rangle ~.
\ea
With the same methods as above we find that the spatial part of the
transition form factor for the Roper resonance is given by
\ba
F(\epsilon) &=&
\langle [56,0^+]_{(1,0)},0| \hat U |[56,0^+]_{(0,0)},0 \rangle
\nonumber\\
&=& \int \mbox{d}\Omega \mbox{d}\Omega^{\prime}
\, Y^{\ast}_{00}(\Omega^{\prime}) \; _u\langle N;R,\Omega^{\prime} |
e^{-i \, \epsilon \hat D_{\lambda,z}} |N;R,\Omega \rangle_c \;
Y_{00}(\Omega)
\nonumber\\
&\rightarrow& -i \, \epsilon \sqrt{\frac{N}{2}} \frac{1-R^2}{1+R^2} \,
\frac{1}{2} \int \mbox{d}(\cos \theta) \, \cos \theta \,
e^{-i \, \epsilon [N R \sqrt{2}/(1+R^2)] \cos \theta}
\nonumber\\
&=& - \frac{1-R^2}{2 R \sqrt{N}} \,
k \beta \, j_1(k \beta) ~.
\ea

The results are summarized in Table~\ref{string1}. For the more
realistic case in which the charge and magnetic moment are not located
at the ends of the string but rather distributed along the string, the
matrix elements are obtained by folding the results of Table~\ref{string1}
with the probability distribution of Eq.~(\ref{prob}). The corresponding
results are presented in Table~\ref{string2}.
For states with angular momentum $L^{\pi}=2^+$ one has to project,
in addition to good angular momentum $L$, also to good permutation
symmetry. We have not carried out the projection
to good permutation symmetry explicitly. It only gives
an overall multiplicative factor and does not change the $k$ dependence
of the matrix elements.

\clearpage
\begin{table}
\centering
\caption[Permutation symmetry]{\small
Transformation properties under $S_3$ of boson creation operators,
generators of the algebra of $U(7)$ and boson-pair
creation operators. Here $l=0,1,2$ and $l^{\prime}=0,2$.
\normalsize}
\label{sthree} \vspace{15pt}
\begin{tabular}{c|c}
\hline
& \\
Operator & $S_3$ \\
& \\
\hline
& \\
$s^{\dagger}$ & $S$ \\
$\hat{n}_s,\hat G^{(l)}_{S}$ & \\
$(b^{\dagger}_{\rho} \times b^{\dagger}_{\rho} + b^{\dagger}_{\lambda}
\times b^{\dagger}_{\lambda})^{(l^{\prime})}$ & \\
& \\
$b_{\rho}^{\dagger}$ & $M_{\rho}$ \\
$\hat D_{\rho},\hat A_{\rho},\hat G^{(l)}_{M_{\rho}}$ & \\
$(b^{\dagger}_{\rho} \times b^{\dagger}_{\lambda} +
b^{\dagger}_{\lambda} \times b^{\dagger}_{\rho})^{(l^{\prime})}$ & \\
& \\
$b_{\lambda}^{\dagger}$ & $M_{\lambda}$ \\
$\hat D_{\lambda},\hat A_{\lambda},\hat G^{(l)}_{M_{\lambda}}$ & \\
$(b^{\dagger}_{\rho} \times b^{\dagger}_{\rho} - b^{\dagger}_{\lambda}
\times b^{\dagger}_{\lambda})^{(l^{\prime})}$ & \\
& \\
$\hat G^{(l)}_A$ & $A$ \\
$(b^{\dagger}_{\rho} \times b^{\dagger}_{\lambda} -
b^{\dagger}_{\lambda} \times b^{\dagger}_{\rho})^{(1)}$ & \\
& \\
\hline
\end{tabular}
\end{table}

\clearpage
\begin{table}
\centering
\caption[Nucleon and delta resonances]{\small
Mass spectrum of nonstrange baryon resonances of the nucleon and delta
family in the collective string model. Here $(n_1,n_2)=(n_u,n_v+n_w)$
denote the vibrational quantum numbers, $K$ is the projection of the
angular momentum $L$, $\pi$ is the parity, $t$ is the transformation
property under the point group $D_3$, and $S$ denotes the spin.
The masses are given in MeV.
The experimental values are taken from \cite{PDG}.
\normalsize}
\label{masses} \vspace{15pt}
\begin{tabular}{llcrcrccc}
\hline
& & & & & & & & \\
Baryon & Status & Mass & $J^{\pi}$ & $(n_1,n_2)$
& $L^{\pi},K$ & $S$ & $t$ & $M_{\mbox{calc}}$ \\
& & & & & & & & \\
\hline
& & & & & & & & \\
N$(939)P_{11}$ & **** & 939 & ${1\over 2}^{+}$ & (0,0) &
$0^{+},0$ & ${1\over 2}$ & $A_{1}$ & 939 \\
N$(1440)P_{11}$ & **** & 1430-1470 & ${1\over 2}^{+}$ & (1,0) &
$0^{+},0$ & ${1\over 2}$ & $A_{1}$ & 1440 \\
N$(1520)D_{13}$ & **** & 1515-1530 & ${3\over 2}^{-}$ & (0,0) &
$1^{-},1$ & ${1\over 2}$ & $E$ & 1566 \\
N$(1535)S_{11}$ & **** & 1520-1555 & ${1\over 2}^{-}$ & (0,0) &
$1^{-},1$ & ${1\over 2}$ & $E$ & 1566 \\
N$(1650)S_{11}$ & **** & 1640-1680 & ${1\over 2}^{-}$ & (0,0) &
$1^{-},1$ & ${3\over 2}$ & $E$ & 1680 \\
N$(1675)D_{15}$ & **** & 1670-1685 & ${5\over 2}^{-}$ & (0,0) &
$1^{-},1$ & ${3\over 2}$ & $E$ & 1680 \\
N$(1680)F_{15}$ & **** & 1675-1690 & ${5\over 2}^{+}$ & (0,0) &
$2^{+},0$ & ${1\over 2}$ & $A_{1}$ & 1735 \\
N$(1700)D_{13}$ & *** & 1650-1750 & ${3\over 2}^{-}$ & (0,0) &
$1^{-},1$ & ${3\over 2}$ & $E$ & 1680 \\
N$(1710)P_{11}$ & *** & 1680-1740 & ${1\over 2}^{+}$ & (0,1) &
$0^{+},0$ & ${1\over 2}$ & $E$ & 1710 \\
N$(1720)P_{13}$ & **** & 1650-1750 & ${3\over 2}^{+}$ & (0,0) &
$2^{+},0$ & ${1\over 2}$ & $A_{1}$ & 1735 \\
N$(2190)G_{17}$ & **** & 2100-2200 & ${7\over 2}^{-}$ & (0,0) &
$3^{-},1$ & ${1\over 2}$ & $E$ & 2140 \\
N$(2220)H_{19}$ & **** & 2180-2310 & ${9\over 2}^{+}$ & (0,0) &
$4^{+},0$ & ${1\over 2}$ & $A_{1}$ & 2267 \\
N$(2250)G_{19}$ & **** & 2170-2310 & ${9\over 2}^{-}$ & (0,0) &
$3^{-},1$ & ${3\over 2}$ & $E$ & 2225 \\
N$(2600)I_{1,11}$ & *** & 2550-2750 & ${11\over 2}^{-}$ & (0,0) &
$5^{-},1$ & ${1\over 2}$ & $E$ & 2590 \\
& & & & & & & & \\
\hline
& & & & & & & & \\
$\Delta(1232)P_{33}$ & **** & 1230-1234 & ${3\over 2}^{+}$ & (0,0) &
$0^{+},0$ & ${3\over 2}$ & $A_{1}$ & 1232 \\
$\Delta(1600)P_{33}$ & *** & 1550-1700 & ${3\over 2}^{+}$ & (1,0) &
$0^{+},0$ & ${3\over 2}$ & $A_{1}$ & 1646 \\
$\Delta(1620)S_{31}$ & **** & 1615-1675 & ${1\over 2}^{-}$ & (0,0) &
$1^{-},1$ & ${1\over 2}$ & $E$ & 1649 \\
$\Delta(1700)D_{33}$ & **** & 1670-1770 & ${3\over 2}^{-}$ & (0,0) &
$1^{-},1$ & ${1\over 2}$ & $E$ & 1649 \\
$\Delta(1900)S_{31}$ & *** & 1850-1950 & ${1\over 2}^{-}$ & (1,0) &
$1^{-},1$ & ${1\over 2}$ & $E$ & 1977 \\
$\Delta(1905)F_{35}$ & **** & 1870-1920 & ${5\over 2}^{+}$ & (0,0) &
$2^{+},0$ & ${3\over 2}$ & $A_{1}$ & 1909 \\
$\Delta(1910)P_{31}$ & **** & 1870-1920 & ${1\over 2}^{+}$ & (0,0) &
$2^{+},0$ & ${3\over 2}$ & $A_{1}$ & 1909 \\
$\Delta(1920)P_{33}$ & *** & 1900-1970 & ${3\over 2}^{+}$ & (0,0) &
$2^{+},0$ & ${3\over 2}$ & $A_{1}$ & 1909 \\
$\Delta(1930)D_{35}$ & *** & 1920-1970 & ${5\over 2}^{-}$ & (0,0) &
$2^{-},1$ & ${1\over 2}$ & $E$ & 1945 \\
$\Delta(1950)F_{37}$ & **** & 1940-1960 & ${7\over 2}^{+}$ & (0,0) &
$2^{+},0$ & ${3\over 2}$ & $A_{1}$ & 1909 \\
$\Delta(2420)H_{3,11}$ & **** & 2300-2500 & ${11\over 2}^{+}$ & (0,0) &
$4^{+},0$ & ${3\over 2}$ & $A_{1}$ & 2403 \\
& & & & & & & & \\
\hline
& & & & & & & &
\end{tabular}
\end{table}

\clearpage
\begin{table}
\centering
\caption[Missing nucleon resonances]{\small
All calculated nucleon resonances (in MeV) below 2 GeV in the collective
string model. Tentative assignments of 1 and 2 star resonances are shown
in brackets.
\normalsize}
\label{missingn} \vspace{15pt}
\begin{tabular}{ccccrc}
\hline
& & & & & \\
State & ($n_1,n_2$) & $K$ & $M_{\mbox{calc}}$ & Baryon \\
& & & & & \\
\hline
& & & & & \\
$^{2}8_{1/2}[56,0^+]$ & (0,0) & 0 &  939 & N$(939)P_{11}$ \\
$^{2}8_{1/2}[70,1^-]$ & (0,0) & 1 & 1566 & N$(1535)S_{11}$ \\
$^{2}8_{3/2}[70,1^-]$ & (0,0) & 1 & 1566 & N$(1520)D_{13}$ \\
$^{4}8_{1/2}[70,1^-]$ & (0,0) & 1 & 1680 & N$(1650)S_{11}$ \\
$^{4}8_{3/2}[70,1^-]$ & (0,0) & 1 & 1680 & N$(1700)D_{13}$ \\
$^{4}8_{5/2}[70,1^-]$ & (0,0) & 1 & 1680 & N$(1675)D_{15}$ \\
$^{2}8_{1/2}[20,1^+]$ & (0,0) & 0 & 1720 & \\
$^{2}8_{3/2}[20,1^+]$ & (0,0) & 0 & 1720 & \\
$^{2}8_{3/2}[56,2^+]$ & (0,0) & 0 & 1735 & N$(1720)P_{13}$ \\
$^{2}8_{5/2}[56,2^+]$ & (0,0) & 0 & 1735 & N$(1680)F_{15}$ \\
$^{2}8_{3/2}[70,2^-]$ & (0,0) & 1 & 1875 & \\
$^{2}8_{5/2}[70,2^-]$ & (0,0) & 1 & 1875 & \\
$^{2}8_{3/2}[70,2^+]$ & (0,0) & 2 & 1875 & [N$(1900)P_{13}$] \\
$^{2}8_{5/2}[70,2^+]$ & (0,0) & 2 & 1875 & [N$(2000)F_{15}$] \\
$^{4}8_{1/2}[70,2^-]$ & (0,0) & 1 & 1972 & \\
$^{4}8_{3/2}[70,2^-]$ & (0,0) & 1 & 1972 & \\
$^{4}8_{5/2}[70,2^-]$ & (0,0) & 1 & 1972 & \\
$^{4}8_{7/2}[70,2^-]$ & (0,0) & 1 & 1972 & \\
$^{4}8_{1/2}[70,2^+]$ & (0,0) & 2 & 1972 & \\
$^{4}8_{3/2}[70,2^+]$ & (0,0) & 2 & 1972 & \\
$^{4}8_{5/2}[70,2^+]$ & (0,0) & 2 & 1972 & \\
$^{4}8_{7/2}[70,2^+]$ & (0,0) & 2 & 1972 & [N$(1990)F_{17}$] \\
& & & & & \\
$^{2}8_{1/2}[56,0^+]$ & (1,0) & 0 & 1440 & N$(1440)P_{11}$ \\
$^{2}8_{1/2}[70,1^-]$ & (1,0) & 1 & 1909 & \\
$^{2}8_{3/2}[70,1^-]$ & (1,0) & 1 & 1909 & \\
& & & & & \\
$^{2}8_{1/2}[70,0^+]$ & (0,1) & 0 & 1710 & N$(1710)P_{11}$ \\
$^{4}8_{3/2}[70,0^+]$ & (0,1) & 0 & 1815 & \\
$^{2}8_{1/2}[56,1^-]$ & (0,1) & 1 & 1866 & \\
$^{2}8_{3/2}[56,1^-]$ & (0,1) & 1 & 1866 & \\
$^{2}8_{1/2}[70,1^+]$ & (0,1) & 0 & 1997 & \\
$^{2}8_{3/2}[70,1^+]$ & (0,1) & 0 & 1997 & \\
$^{2}8_{1/2}[70,1^-]$ & (0,1) & 1 & 1997 & \\
$^{2}8_{3/2}[70,1^-]$ & (0,1) & 1 & 1997 & \\
& & & & & \\
\hline
\end{tabular}
\end{table}

\clearpage
\begin{table}
\centering
\caption[Missing delta resonances]{\small
All calculated delta resonances (in MeV) below 2 GeV in the collective
string model. Tentative assignments of 1 and 2 star resonances are shown
in brackets.
\normalsize}
\label{missingd} \vspace{15pt}
\begin{tabular}{ccccrc}
\hline
& & & & & \\
State & ($n_1,n_2$) & $K$ & $M_{\mbox{calc}}$ & Baryon \\
& & & & & \\
\hline
& & & & & \\
$^{4}10_{3/2}[56,0^+]$ & (0,0) & 0 & 1232 & $\Delta(1232)P_{33}$ \\
$^{2}10_{1/2}[70,1^-]$ & (0,0) & 1 & 1649 & $\Delta(1620)S_{31}$ \\
$^{2}10_{3/2}[70,1^-]$ & (0,0) & 1 & 1649 & $\Delta(1700)D_{33}$ \\
$^{4}10_{1/2}[56,2^+]$ & (0,0) & 0 & 1909 & $\Delta(1910)P_{31}$ \\
$^{4}10_{3/2}[56,2^+]$ & (0,0) & 0 & 1909 & $\Delta(1920)P_{33}$ \\
$^{4}10_{5/2}[56,2^+]$ & (0,0) & 0 & 1909 & $\Delta(1905)F_{35}$ \\
$^{4}10_{7/2}[56,2^+]$ & (0,0) & 0 & 1909 & $\Delta(1950)F_{37}$ \\
$^{2}10_{3/2}[70,2^-]$ & (0,0) & 1 & 1945 & [$\Delta(1940)D_{33}$] \\
$^{2}10_{5/2}[70,2^-]$ & (0,0) & 1 & 1945 & $\Delta(1930)D_{35}$ \\
$^{2}10_{3/2}[70,2^+]$ & (0,0) & 2 & 1945 & \\
$^{2}10_{5/2}[70,2^+]$ & (0,0) & 2 & 1945 & [$\Delta(2000)F_{35}$] \\
& & & & & \\
$^{4}10_{3/2}[56,0^+]$ & (1,0) & 0 & 1646 & $\Delta(1600)P_{33}$ \\
$^{2}10_{1/2}[70,1^-]$ & (1,0) & 1 & 1977 & $\Delta(1900)S_{31}$ \\
$^{2}10_{3/2}[70,1^-]$ & (1,0) & 1 & 1977 & \\
& & & & & \\
$^{2}10_{1/2}[70,0^+]$ & (0,1) & 0 & 1786 & [$\Delta(1750)P_{31}$] \\
& & & & & \\
\hline
\end{tabular}
\end{table}

\clearpage
\begin{table}
\centering
\caption[Spin-flavor coefficients: nucleon resonances]{\small
Spin-flavor coefficients of ${\cal H}_{\mbox{nr}}$ in transverse,
Eq.~(\ref{helampt}), longitudinal, Eq.~(\ref{helampl}), and scalar,
Eq.~(\ref{helamps}), helicity amplitudes for nucleon resonances:
proton- and neutron-target couplings.
\normalsize}
\label{emn} \vspace{15pt}
\begin{tabular}{c|cccccccccc}
\hline
& & & & & & & & & & \\
& \multicolumn{2}{c} {$A^p_{1/2}$} & \multicolumn{2}{c} {$A^p_{3/2}$}
& \multicolumn{2}{c} {$A^n_{1/2}$} & \multicolumn{2}{c} {$A^n_{3/2}$}
& $A_l^p \, (A_s^p)$ & $A_l^n \, (A_s^n)$ \\
State & $\alpha_{1/2}$ & $\beta_{1/2}$ & $\alpha_{3/2}$ & $\beta_{3/2}$
      & $\alpha_{1/2}$ & $\beta_{1/2}$ & $\alpha_{3/2}$ & $\beta_{3/2}$
      & $\gamma \, (\delta)$ & $\gamma \, (\delta)$ \\
& & & & & & & & & & \\
\hline
& & & & & & & & & & \\
$^{2}8_{1/2}[56,0^+]$ & 0 & $\frac{1}{3}$ & 0 & 0
& 0 & $\frac{-2}{9}$ & 0 & 0
& $\frac{1}{3}$ & 0 \\
& & & & & & & & & & \\
$^{2}8_{3/2}[56,2^+]$ & $\frac{-1}{\sqrt{15}}$ &
$\frac{-\sqrt{2}}{3\sqrt{5}}$ & $\frac{1}{3\sqrt{5}}$ & 0
& 0 & $\frac{2\sqrt{2}}{9\sqrt{5}}$ & 0 & 0
& $\frac{-\sqrt{2}}{3\sqrt{5}}$ & 0 \\
& & & & & & & & & & \\
$^{2}8_{5/2}[56,2^+]$ & $\frac{-\sqrt{2}}{3\sqrt{5}}$ &
$\frac{1}{\sqrt{15}}$ & $\frac{-2}{3\sqrt{5}}$ & 0
& 0 & $\frac{-2}{3\sqrt{15}}$ & 0 & 0
& $\frac{1}{\sqrt{15}}$ & 0 \\
& & & & & & & & & & \\
$^{2}8_{1/2}[70,0^+]$ & 0 & $\frac{1}{3\sqrt{2}}$ & 0 & 0
& 0 & $\frac{-1}{9\sqrt{2}}$ & 0 & 0
& $\frac{1}{3\sqrt{2}}$ & $\frac{-1}{3\sqrt{2}}$ \\
& & & & & & & & & & \\
$^{2}8_{1/2}[70,1^-]$ & $\frac{-1}{3\sqrt{3}}$ & $\frac{-1}{3\sqrt{6}}$ &
0 & 0 & $\frac{1}{3\sqrt{3}}$ & $\frac{1}{9\sqrt{6}}$ & 0 & 0
& $\frac{-1}{3\sqrt{6}}$ & $\frac{1}{3\sqrt{6}}$ \\
& & & & & & & & & & \\
$^{2}8_{3/2}[70,1^-]$ & $\frac{-1}{3\sqrt{6}}$ & $\frac{1}{3\sqrt{3}}$ &
$\frac{-1}{3\sqrt{2}}$ & 0 & $\frac{1}{3\sqrt{6}}$ &
$\frac{-1}{9\sqrt{3}}$ & $\frac{1}{3\sqrt{2}}$ & 0
& $\frac{1}{3\sqrt{3}}$ & $\frac{-1}{3\sqrt{3}}$ \\
& & & & & & & & & & \\
$^{2}8_{3/2}[70,2^+]$ & $\frac{-1}{\sqrt{30}}$ &
$\frac{-1}{3\sqrt{5}}$ & $\frac{1}{3\sqrt{10}}$ & 0
& $\frac{1}{\sqrt{30}}$ &
$\frac{1}{9\sqrt{5}}$ & $\frac{-1}{3\sqrt{10}}$ & 0
& $\frac{-1}{3\sqrt{5}}$ & $\frac{1}{3\sqrt{5}}$ \\
& & & & & & & & & & \\
$^{2}8_{5/2}[70,2^+]$ & $\frac{-1}{3\sqrt{5}}$ &
$\frac{1}{\sqrt{30}}$ & $\frac{-\sqrt{2}}{3\sqrt{5}}$ & 0
& $\frac{1}{3\sqrt{5}}$ &
$\frac{-1}{3\sqrt{30}}$ & $\frac{\sqrt{2}}{3\sqrt{5}}$ & 0
& $\frac{1}{\sqrt{30}}$ & $\frac{-1}{\sqrt{30}}$ \\
& & & & & & & & & & \\
$^{4}8_{3/2}[70,0^+]$ & 0 & 0 & 0 & 0
& 0 & $\frac{1}{9\sqrt{2}}$ & 0 & $\frac{1}{3\sqrt{6}}$
& 0 & 0 \\
& & & & & & & & & & \\
$^{4}8_{1/2}[70,1^-]$ & 0 & 0 & 0 & 0
& 0 & $\frac{-1}{9\sqrt{6}}$ & 0 & 0 & 0 & 0 \\
& & & & & & & & & & \\
$^{4}8_{3/2}[70,1^-]$ & 0 & 0 & 0 & 0
& 0 & $\frac{-1}{9\sqrt{30}}$ & 0 & $\frac{-1}{3\sqrt{10}}$
& 0 & 0 \\
& & & & & & & & & & \\
$^{4}8_{5/2}[70,1^-]$ & 0 & 0 & 0 & 0
& 0 & $\frac{1}{3\sqrt{30}}$ & 0 & $\frac{1}{3\sqrt{15}}$
& 0 & 0 \\
& & & & & & & & & & \\
$^{4}8_{1/2}[70,2^+]$ & 0 & 0 & 0 & 0
& 0 & $\frac{1}{9\sqrt{10}}$ & 0 & 0 & 0 & 0 \\
& & & & & & & & & & \\
$^{4}8_{3/2}[70,2^+]$ & 0 & 0 & 0 & 0
& 0 & $\frac{-1}{9\sqrt{10}}$ & 0 & $\frac{1}{3\sqrt{30}}$
& 0 & 0 \\
& & & & & & & & & & \\
$^{4}8_{5/2}[70,2^+]$ & 0 & 0 & 0 & 0
& 0 & $\frac{-1}{3\sqrt{210}}$ & 0 & $\frac{-1}{\sqrt{105}}$
& 0 & 0 \\
& & & & & & & & & & \\
$^{4}8_{7/2}[70,2^+]$ & 0 & 0 & 0 & 0
& 0 & $\frac{1}{3\sqrt{35}}$ & 0 & $\frac{1}{3\sqrt{21}}$ & 0 & 0 \\
& & & & & & & & & & \\
$^{2}8_{1/2}[20,1^+]$ & 0 & 0 & 0 & 0 & 0 & 0 & 0 & 0 & 0 & 0 \\
& & & & & & & & & & \\
$^{2}8_{3/2}[20,1^+]$ & 0 & 0 & 0 & 0 & 0 & 0 & 0 & 0 & 0 & 0 \\
& & & & & & & & & & \\
\hline
\end{tabular}
\end{table}

\clearpage
\begin{table}
\centering
\caption[Spin-flavor coefficients: delta resonances]{\small
Spin-flavor coefficients of ${\cal H}_{\mbox{nr}}$ in transverse,
Eq.~(\ref{helampt}), longitudinal, Eq.~(\ref{helampl}), and scalar,
Eq.~(\ref{helamps}), helicity amplitudes for delta resonances.
\normalsize}
\label{emd} \vspace{15pt}
\begin{tabular}{c|ccccc}
\hline
& & & & & \\
& \multicolumn{2}{c} {$A^{p,n}_{1/2}$}
& \multicolumn{2}{c} {$A^{p,n}_{3/2}$}
& $A_l^{p,n} \, (A_s^{p,n})$ \\
State & $\alpha_{1/2}$ & $\beta_{1/2}$ & $\alpha_{3/2}$
& $\beta_{3/2}$ & $\gamma  \, (\delta)$ \\
& & & & & \\
\hline
& & & & & \\
$^{4}10_{3/2}[56,0^+]$ & 0 & $\frac{-\sqrt{2}}{9}$ &
0 & $\frac{-\sqrt{2}}{3\sqrt{3}}$ & 0 \\
& & & & & \\
$^{4}10_{1/2}[56,2^+]$ & 0 & $\frac{-\sqrt{2}}{9\sqrt{5}}$ & 0 & 0 & 0 \\
& & & & & \\
$^{4}10_{3/2}[56,2^+]$ & 0 & $\frac{\sqrt{2}}{9\sqrt{5}}$ &
0 & $\frac{-\sqrt{2}}{3\sqrt{15}}$ & 0 \\
& & & & & \\
$^{4}10_{5/2}[56,2^+]$ & 0 & $\frac{\sqrt{2}}{3\sqrt{105}}$ &
0 & $\frac{2}{\sqrt{105}}$ & 0 \\
& & & & & \\
$^{4}10_{7/2}[56,2^+]$ & 0 & $\frac{-2}{3\sqrt{35}}$ &
0 & $\frac{-2}{3\sqrt{21}}$ & 0 \\
& & & & & \\
$^{2}10_{1/2}[70,0^+]$ & 0 & $\frac{1}{9\sqrt{2}}$ & 0 & 0
& $\frac{-1}{3\sqrt{2}}$ \\
& & & & & \\
$^{2}10_{1/2}[70,1^-]$ & $\frac{1}{3\sqrt{3}}$ &
$\frac{-1}{9\sqrt{6}}$ & 0 & 0 & $\frac{1}{3\sqrt{6}}$ \\
& & & & & \\
$^{2}10_{3/2}[70,1^-]$ & $\frac{1}{3\sqrt{6}}$ &
$\frac{1}{9\sqrt{3}}$ & $\frac{1}{3\sqrt{2}}$ & 0
& $\frac{-1}{3\sqrt{3}}$ \\
& & & & & \\
$^{2}10_{3/2}[70,2^+]$ & $\frac{1}{\sqrt{30}}$ &
$\frac{-1}{9\sqrt{5}}$ & $\frac{-1}{3\sqrt{10}}$ & 0
& $\frac{1}{3\sqrt{5}}$ \\
& & & & & \\
$^{2}10_{5/2}[70,2^+]$ & $\frac{1}{3\sqrt{5}}$ &
$\frac{1}{3\sqrt{30}}$ & $\frac{\sqrt{2}}{3\sqrt{5}}$ & 0
& $\frac{-1}{\sqrt{30}}$ \\
& & & & & \\
\hline
\end{tabular}
\end{table}

\clearpage
\begin{table}
\centering
\caption[Form factors in the harmonic oscillator model]{\small
Analytic expressions of the matrix elements of the transition operators of
Eq.~(\ref{emop}) in the harmonic oscillator limit of $U(7)$
for $N \rightarrow \infty$. The initial state is $[56,0^+]_0$.
\normalsize}
\label{ho} \vspace{15pt}
\begin{tabular}{cccc}
\hline
& & & \\
Final state & $\langle f | \hat U | i \rangle$
& $\langle f | \hat T_{z} | i \rangle/m_q k_0 \beta$
& $\langle f | \hat T_{\pm} | i \rangle/m_q k_0 \beta$ \\
& & & \\
\hline
& & & \\
$[56,0^+]_0$ & $\mbox{e}^{-k^2 \beta^2/6}$
& $\frac{1}{3} k \beta \, \mbox{e}^{-k^2 \beta^2/6}$ & 0 \\
& & & \\
$[70,1^-]_1$ & $-i \frac{1}{\sqrt{3}} k \beta
\, \mbox{e}^{-k^2 \beta^2/6}$
& $i \frac{1}{\sqrt{3}}(1-\frac{k^2 \beta^2}{3})
\, \mbox{e}^{-k^2 \beta^2/6}$
& $\mp i \sqrt{\frac{2}{3}} \, \mbox{e}^{-k^2 \beta^2/6}$ \\
& & & \\
$[56,0^+]_2$
& $\frac{1}{6\sqrt{3}} k^2 \beta^2 \, \mbox{e}^{-k^2 \beta^2/6}$
& $-\frac{1}{3\sqrt{3}} k \beta(1-\frac{k^2 \beta^2}{6})
\, \mbox{e}^{-k^2 \beta^2/6}$ & 0 \\
& & & \\
$[70,0^+]_2$
& $-\frac{1}{6\sqrt{3}} k^2 \beta^2 \, \mbox{e}^{-k^2 \beta^2/6}$
& $\frac{1}{3\sqrt{3}} k \beta(1-\frac{k^2 \beta^2}{6})
\, \mbox{e}^{-k^2 \beta^2/6}$ & 0 \\
& & & \\
$[56,2^+]_2$
& $-\frac{1}{3\sqrt{6}} k^2 \beta^2 \, \mbox{e}^{-k^2 \beta^2/6}$
& $\frac{2}{3\sqrt{6}} k \beta(1-\frac{k^2 \beta^2}{6})
\, \mbox{e}^{-k^2 \beta^2/6}$
& $\mp \frac{1}{3} k \beta \, \mbox{e}^{-k^2 \beta^2/6}$ \\
& & & \\
$[70,2^+]_2$
& $\frac{1}{3\sqrt{6}} k^2 \beta^2 \, \mbox{e}^{-k^2 \beta^2/6}$
& $-\frac{2}{3\sqrt{6}} k \beta(1-\frac{k^2 \beta^2}{6})
\, \mbox{e}^{-k^2 \beta^2/6}$
& $\pm \frac{1}{3} k \beta \, \mbox{e}^{-k^2 \beta^2/6}$ \\
& & & \\
\hline
& & & \\
\end{tabular}
\end{table}

\clearpage
\begin{table}
\centering
\caption[Form factors in the end string model]{\small
Analytic expressions of the matrix elements of the transition operators of
Eq.~(\ref{emop}) in the end string model
for $N \rightarrow \infty$. The initial state is $[56,0^+]_{(0,0)}$.
\normalsize}
\label{string1} \vspace{15pt}
\begin{tabular}{cccc}
\hline
& & & \\
Final state & $\langle f | \hat U | i \rangle$
& $\langle f | \hat T_{z} | i \rangle/m_q k_0 \beta$
& $\langle f | \hat T_{\pm} | i \rangle/m_q k_0 \beta$ \\
& & & \\
\hline
& & & \\
$[56,0^+]_{(0,0)}$ & $j_0(k \beta)$ & $j_1(k \beta)$ & 0 \\
& & & \\
$[70,1^-]_{(0,0)}$ & $-i \, \sqrt{3} \, j_1(k \beta)$
& $i \frac{1}{\sqrt{3}} [j_0(k \beta)-2j_2(k \beta)]$
& $\mp i \, \sqrt{\frac{2}{3}} [j_0(k \beta)+ j_2(k \beta)]$ \\
& & & \\
$[56,0^+]_{(1,0)}$
& $- \frac{1-R^2}{2R \sqrt{N}} \, k \beta \, j_1(k \beta)$
& $\frac{1-R^2}{6R \sqrt{N}} \, k \beta [2j_0(k \beta)-j_2(k \beta)]$
& 0 \\
& & & \\
$[70,0^+]_{(0,1)}$
& $- \frac{1}{2} \sqrt{\frac{1+R^2}{NR^2}} \, k \beta \, j_1(k \beta)$
& $\frac{1}{6} \sqrt{\frac{1+R^2}{NR^2}} \,
k \beta [2j_0(k \beta)-j_2(k \beta)]$ & 0 \\
& & & \\
$[56,2^+]_{(0,0)}^{a}$ & $-\sqrt{5} \, j_2(k \beta)$
& $\frac{1}{\sqrt{5}} [2j_1(k \beta)-3j_3(k \beta)]$
& $\mp \sqrt{\frac{6}{5}} [j_1(k \beta)+j_3(k \beta)]$ \\
& & & \\
$[70,2^+]_{(0,0)}^{a}$ & $-\sqrt{5} \, j_2(k \beta)$
& $\frac{1}{\sqrt{5}} [2j_1(k \beta)-3j_3(k \beta)]$
& $\mp \sqrt{\frac{6}{5}} [j_1(k \beta)+j_3(k \beta)]$ \\
& & & \\
\hline
& & & \\
\multicolumn{4}{l} {$^{a}$ Up to an overall constant.} \\
\end{tabular}
\end{table}

\clearpage
\begin{table}
\centering
\caption[Form factors in the distributed string model]{\small
Analytic expressions of the matrix elements of the transition operators of
Eq.~(\ref{emop}) in the distributed string model
for $N \rightarrow \infty$. $H(x)=\arctan x - x/(1+x^2)$.
The initial state is $[56,0^+]_{(0,0)}$.
\normalsize}
\label{string2} \vspace{15pt}
\begin{tabular}{cccc}
\hline
& & & \\
Final state & $\langle f | \hat U | i \rangle$
& $\langle f | \hat T_{z}   | i \rangle/m_q k_0 a$
& $\langle f | \hat T_{\pm} | i \rangle/m_q k_0 a$ \\
& & & \\
\hline
& & & \\
$[56,0^+]_{(0,0)}$
& $\frac{1}{(1+k^2a^2)^2}$ & $\frac{4ka}{(1+k^2a^2)^3}$ & 0  \\
& & & \\
$[70,1^-]_{(0,0)}$ & $-i \, \sqrt{3} \, \frac{ka}{(1+k^2a^2)^2}$
& $i \, \sqrt{3} \, \frac{1-3k^2a^2}{(1+k^2a^2)^3}$
& $\mp i \, \sqrt{6} \, \frac{1}{(1+k^2a^2)^2}$ \\
& & & \\
$[56,0^+]_{(1,0)}$
& $- \frac{1-R^2}{R \sqrt{N}} \frac{2k^2a^2}{(1+k^2a^2)^3}$
& $  \frac{1-R^2}{R \sqrt{N}} \frac{4ka(1-2k^2a^2)}{(1+k^2a^2)^4}$ & 0 \\
& & & \\
$[70,0^+]_{(0,1)}$
& $- \sqrt{\frac{1+R^2}{NR^2}} \frac{2k^2a^2}{(1+k^2a^2)^3}$
& $  \sqrt{\frac{1+R^2}{NR^2}} \frac{4ka(1-2k^2a^2)}{(1+k^2a^2)^4}$ & 0 \\
& & & \\
$[56,2^+]_{(0,0)}^{a}$
& $-\sqrt{5}\left[ \frac{-1}{(1+k^2a^2)^2} \right.$
& $\sqrt{5}\left[ \frac{3+7k^2a^2}{ka(1+k^2a^2)^3} \right.$
& $\mp \sqrt{30}\left[ \frac{-1}{ka(1+k^2a^2)^2} \right.$ \\
& $\left. \hspace{1cm} + \frac{3}{2k^3a^3} H(ka) \right]$
& $\left. \hspace{1cm} - \frac{9}{2k^4a^4} H(ka) \right]$
& $\left. \hspace{1cm} + \frac{3}{2k^4a^4} H(ka) \right]$ \\
& & & \\
$[70,2^+]_{(0,0)}^{a}$
& $-\sqrt{5}\left[ \frac{-1}{(1+k^2a^2)^2} \right.$
& $\sqrt{5}\left[ \frac{3+7k^2a^2}{ka(1+k^2a^2)^3} \right.$
& $\mp \sqrt{30}\left[ \frac{-1}{ka(1+k^2a^2)^2} \right.$ \\
& $\left. \hspace{1cm} + \frac{3}{2k^3a^3} H(ka) \right]$
& $\left. \hspace{1cm} - \frac{9}{2k^4a^4} H(ka) \right]$
& $\left. \hspace{1cm} + \frac{3}{2k^4a^4} H(ka) \right]$ \\
& & & \\
\hline
& & & \\
\multicolumn{4}{l} {$^{a}$ Up to an overall constant.} \\
\end{tabular}
\end{table}

\clearpage
\begin{table}
\centering
\caption[Photocouplings for nucleon resonances]{\small
Helicity amplitudes $A^{p,n}_{\mu}$ in 10$^{-3}$ GeV$^{-1/2}$ for nucleon
resonances calculated in the Breit frame for $R^2=0.0$ (harmonic
oscillator) (1), $R^2=0.5$ and $R^2=1.0$ (distributed string), (2)
and (3), respectively. The calculations are done in a model space with
$n_{\rho}+n_{\lambda} \leq N=20$. The quark mass $m_q$ is
$M_p/2.793$, which corresponds to $g=1$ and $\mu=\mu_p=0.13$ GeV$^{-1}$.
The size parameters are obtained from the r.m.s. radius and are given
by $\beta=0.855$ fm, $a=0.242$ fm and $a=0.248$ fm, respectively.
We have suppressed a factor of $+i$ for the transitions to the negative
parity states. The data are taken from \cite{PDG}.
\normalsize}
\label{nphoto} \vspace{15pt}
\begin{tabular}{lcrrrrc}
\hline
& & & & & & \\
Resonance & State & $\mu$ & \multicolumn{3}{c} {$A_{\mu}^{p,n}$(th)}
& $A_{\mu}^{p,n}$(exp) \\
& & & (1) & (2) & (3) & \\
& & & & & & \\
\hline
& & & & & & \\
N$(1440)P_{11}$ & $^{2}8_{1/2}[56,0^+]$
  & $p,1/2$ & $+67$ & $+12$ & $+0$ & $-68 \pm 5$ \\
& & $n,1/2$ & $-45$ & $- 8$ & $-0$ & $+39 \pm 15$ \\
N$(1520)D_{13}$ & $^{2}8_{3/2}[70,1^-]$
  & $p,1/2$ & $- 43$ & $- 42$ & $- 43$ & $- 23 \pm  9$ \\
& & $n,1/2$ & $- 27$ & $- 27$ & $- 27$ & $- 64 \pm  8$ \\
& & $p,3/2$ & $+108$ & $+107$ & $+109$ & $+163 \pm  8$ \\
& & $n,3/2$ & $-108$ & $-107$ & $-109$ & $-141 \pm 11$ \\
N$(1535)S_{11}$ & $^{2}8_{1/2}[70,1^-]$
  & $p,1/2$ & $+158$ & $+160$ & $+162$ & $+74 \pm 11$ \\
& &         & $+125$ & $+126$ & $+127$$^{a}$ & \\
& & $n,1/2$ & $-109$ & $-111$ & $-112$ & $-72 \pm 25$ \\
& &         & $-101$ & $-102$ & $-103$$^{a}$ & \\
N$(1650)S_{11}$ & $^{4}8_{1/2}[70,1^-]$
  & $p,1/2$ & 0 & 0 & 0 & $+48 \pm 16$ \\
& &         & $+75$ & $+91$ & $+91$$^{a}$ & \\
& & $n,1/2$ & $+21$ & $+25$ & $+25$ & $-17 \pm 37$ \\
& &         & $-33$ & $-40$ & $-41$$^{a}$ & \\
N$(1675)D_{15}$ & $^{4}8_{5/2}[70,1^-]$
  & $p,1/2$ & 0 & 0 & 0 & $+19 \pm 12$ \\
& & $n,1/2$ & $-26$ & $-33$ & $-33$ & $-47 \pm 23$ \\
& & $p,3/2$ & 0 & 0 & 0 & $+19 \pm 12$ \\
& & $n,3/2$ & $-37$ & $-47$ & $-47$ & $-69 \pm 19$ \\
N$(1680)F_{15}$ & $^{2}8_{5/2}[56,2^+]$
  & $p,1/2$ & $-6$ & $-4$ & $-4$ & $ -17 \pm 10$ \\
& & $n,1/2$ & $+ 55$ & $+40$ & $+40$ & $+ 31 \pm 13$ \\
& & $p,3/2$ & $+109$ & $+81$ & $+80$ & $+127 \pm 12$ \\
& & $n,3/2$ & 0 & 0 & 0 & $ -30 \pm 14$ \\
N$(1700)D_{13}$ & $^{4}8_{3/2}[70,1^-]$
  & $p,1/2$ & 0 & 0 & 0 & $-22 \pm 13$ \\
& & $n,1/2$ & $+ 8$ & $+11$ & $+11$ & $  0 \pm 56$ \\
& & $p,3/2$ & 0 & 0 & 0 & $  0 \pm 19$ \\
& & $n,3/2$ & $+43$ & $+57$ & $+57$ & $ -2 \pm 44$ \\
N$(1710)P_{11}$ & $^{2}8_{1/2}[70,0^+]$
  & $p,1/2$ & $-52$ & $-27$ & $-22$ & $+5 \pm 16$ \\
& & $n,1/2$ & $+17$ & $+ 9$ & $+ 7$ & $-5 \pm 23$ \\
N$(1720)P_{13}$ & $^{2}8_{3/2}[56,2^+]$
  & $p,1/2$ & $+151$ & $+119$ & $+118$ & $+52 \pm 39$ \\
& & $n,1/2$ & $ -43$ & $ -34$ & $ -33$ & $ -2 \pm 26$ \\
& & $p,3/2$ & $ -50$ & $ -39$ & $ -39$ & $-35 \pm 24$ \\
& & $n,3/2$ & 0 & 0 & 0 & $-43 \pm 94$ \\
N$(1990)F_{17}$ & $^{4}8_{7/2}[70,2^+]$
  & $p,1/2$ & 0 & 0 & 0 & $ +24 \pm 30$ \\
& & $n,1/2$ & $+ 9$ & $+22$ & $+23$ & $ -49 \pm 45$ \\
& & $p,3/2$ & 0 & 0 & 0 & $ +31 \pm 55$ \\
& & $n,3/2$ & $+12$ & $+28$ & $+29$ & $-122 \pm 55$ \\
& & & & & & \\
\hline
& & & & & & \\
\multicolumn{7}{l} {$^{a}$ The results in this line are obtained by
introducing a mixing angle as in Eq.~(\ref{mix}).} \\
\end{tabular}
\end{table}

\clearpage
\begin{table}
\centering
\caption[Photocouplings for delta resonances]{\small
Same as Table~\ref{nphoto}, but for delta resonances.
\normalsize}
\label{dphoto} \vspace{15pt}
\begin{tabular}{lccrrrc}
\hline
& & & & & & \\
Resonance & State & $\mu$ & \multicolumn{3}{c} {$A_{\mu}^{p,n}$(th)}
& $A_{\mu}^{p,n}$(exp) \\
& & & (1) & (2) & (3) & \\
& & & & & & \\
\hline
& & & & & & \\
$\Delta(1232)P_{33}$ & $^{4}10_{3/2}[56,0^+]$
  & $1/2$ & $- 90$ & $- 91$ & $- 91$ & $-141 \pm  5$ \\
& & $3/2$ & $-155$ & $-158$ & $-157$ & $-258 \pm 12$ \\
$\Delta(1600)P_{33}$ & $^{4}10_{3/2}[56,0^+]$
  & $1/2$ & $-37$ & $- 7$ & $+0$ & $-20 \pm 29$ \\
& & $3/2$ & $-64$ & $-12$ & $+0$ & $+ 1 \pm 22$ \\
$\Delta(1620)S_{31}$ & $^{2}10_{1/2}[70,1^-]$
  & $1/2$ & $-44$ & $-51$ & $-51$ & $+19 \pm 16$ \\
& & & & & & $+30 \pm 10$ \\
$\Delta(1700)D_{33}$ & $^{2}10_{3/2}[70,1^-]$
  & $1/2$ & $-62$ & $-82$ & $-82$ & $+116 \pm 17$ \\
& & $3/2$ & $-62$ & $-83$ & $-82$ & $+ 77 \pm 28$ \\
$\Delta(1900)S_{31}$ & $^{2}10_{1/2}[70,1^-]$
  & $1/2$ & $-2$ & $-1$ & $+0$ & $+10 \pm ?$ \\
$\Delta(1905)F_{35}$ & $^{4}10_{5/2}[56,2^+]$
  & $1/2$ & $-10$ & $-12$ & $-11$ & $+27 \pm 13$ \\
& & $3/2$ & $-41$ & $-49$ & $-49$ & $-47 \pm 19$ \\
$\Delta(1910)P_{31}$ & $^{4}10_{1/2}[56,2^+]$
  & $1/2$ & $+15$ & $+18$ & $+17$ & $-12 \pm 30$ \\
$\Delta(1920)P_{33}$ & $^{4}10_{3/2}[56,2^+]$
  & $1/2$ & $-14$ & $-18$ & $-17$ & $+40 \pm ?$ \\
& & $3/2$ & $+25$ & $+31$ & $+30$ & $+23 \pm ?$ \\
$\Delta(1930)D_{35}$ & $^{2}10_{5/2}[70,2^-]$
  & $1/2$ & 0 & 0 & 0 & $-30 \pm 40$ \\
& & $3/2$ & 0 & 0 & 0 & $-10 \pm 35$ \\
$\Delta(1950)F_{37}$ & $^{4}10_{7/2}[56,2^+]$
  & $1/2$ & $+21$ & $+29$ & $+28$ & $-73 \pm 14$ \\
& & $3/2$ & $+27$ & $+37$ & $+36$ & $-90 \pm 13$ \\
& & & & & & \\
\hline
\end{tabular}
\end{table}

\clearpage
\begin{figure}
\vspace{0.5cm}
\caption[Geometric structure of baryons]{
Collective model of baryons and its idealized string configuration
(the charge distribution of the proton is shown as an example).}
\label{geometry}
\end{figure}

\begin{figure}
\vspace{0.5cm}
\caption[Excitations of harmonic oscillator]{
Schematic representation of the spectrum of the harmonic oscillator quark
model with three identical constituents. The excitations are labeled by
$n=n_{\rho}+n_{\lambda}$ and $L^{\pi}_t$, where $\pi$ denotes the parity
and $t$ the transformation property under $S_3$ (the equivalent label
of $D_3$ is used). Each $E$ state is doubly degenerate.}
\label{harmosc}
\end{figure}

\begin{figure}
\vspace{0.5cm}
\caption[Fundamental vibrations]{
Vibrations of the string-like configuration of Figure~\ref{geometry}.}
\label{vibrations}
\end{figure}

\begin{figure}
\vspace{0.5cm}
\caption[Excitations of collective string]{
Schematic representation of the vibrational and rotational excitations
of the string-like configuration of Figure~\ref{geometry} with three
identical constituent parts. The vibrational excitations are labeled by
$(n_u,n_v+n_w)$ and the rotational levels by $K$, $L^{\pi}_t$,
where $K$ is the projection of the angular momentum $L$, $\pi$ denotes
the parity and $t$ the overall (vibrational plus rotational) transformation
property under the point group $D_3$. Each $E$ state is doubly degenerate.}
\label{string}
\end{figure}

\begin{figure}
\vspace{0.5cm}
\caption[Regge trajectories]{
Plot of $M^2$ versus $L$ for a selected number of nucleon resonances.
The lines represent the fit with $\alpha=1.064$ GeV$^2$.}
\label{regge}
\end{figure}

\begin{figure}
\vspace{0.5cm}
\caption[Signature dependence of baryons]{
Mass spectrum of some nucleon resonances. Solid lines: collective
string model, Eq.~(\ref{massformula}). Dotted lines: relativized
quark model \cite{CI}. Circles and vertical lines: experimental masses
and their uncertainties, taken from \cite{PDG}.}
\label{signature}
\end{figure}

\begin{figure}
\vspace{0.5cm}
\caption[Proton form factor]{
Comparison between the experimental proton electric form factor
$G_E^p$ and the calculations with $R^2=0$ (harmonic oscillator,
dotted line), $R^2=0.5$ and 1.0 (distributed string, dashed and
solid line). The experimental data, taken from a compilation in
\cite{IJL}, and the calculations are divided by the dipole form
factor, $F_D=1/(1+Q^2/0.71)^2$.}
\label{protonff}
\end{figure}

\begin{figure}
\vspace{0.5cm}
\caption[Delta form factor]{
Comparison between the experimental magnetic transition form factor for
$\Delta(1232)P_{33}$ and the calculations with $R^2=0$ (harmonic
oscillator, dotted line), $R^2=0.5$ and 1.0 (distributed string, dashed
and solid line). The experimental data, taken from the compilation in
\cite{Burkert}, and the calculations are divided by $3F_D$.}
\label{deltaff}
\end{figure}

\begin{figure}
\vspace{0.5cm}
\caption[Form factors: N$(1520)D_{13}$]{
Comparison between the experimental transition form factor for the
N$(1520)D_{13}$ resonance and the calculations with $R^2=0$ (harmonic
oscillator, dotted line) and $R^2=1.0$ (distributed string, solid line).
The experimental data are taken from the compilation in \cite{Burkert2}.}
\label{d13}
\end{figure}

\begin{figure}
\vspace{0.5cm}
\caption[Form factors: N$(1535)S_{11}$]{
Same as Figure~\ref{d13}, but for the N$(1535)S_{11}$ resonance.
Two calculations are shown, one with no mixing $\theta=0^{\circ}$,
and one with a mixing angle $\theta=-38^{\circ}$ (see Eq.~(\ref{mix})).
The experimental data are taken from the compilation in \cite{Burkert}.}
\label{s11(1535)}
\end{figure}

\begin{figure}
\vspace{0.5cm}
\caption[Form factors: N$(1650)S_{11}$]{
Same Figure~\ref{s11(1535)}, but for the N$(1650)S_{11}$ resonance.
The experimental data are taken from the compilation in \cite{Burkert}.}
\label{s11(1650)}
\end{figure}

\begin{figure}
\vspace{0.5cm}
\caption[Form factors: N$(1680)F_{15}$]{
Same as Figure~\ref{d13}, but for the N$(1680)F_{15}$ resonance.
The experimental data are taken from the compilation in \cite{Burkert2}.}
\label{f15}
\end{figure}

\begin{figure}
\vspace{0.5cm}
\caption[Helicity asymmetry: N$(1520)D_{13}$]{
Comparison between the experimental proton helicity asymmetry for the
N$(1520)D_{13}$ resonance and the calculations with $R^2=0$ (harmonic
oscillator, dotted line) and $R^2=1.0$ (distributed string, solid line).
The experimental data are taken from the compilation in \cite{Burkert}.}
\label{asymd13}
\end{figure}

\begin{figure}
\vspace{0.5cm}
\caption[Helicity asymmetry: N$(1680)F_{15}$]{
Same as Figure~\ref{asymd13}, but for the N$(1680)F_{15}$ resonance.
The experimental data are taken from the compilation in \cite{Burkert}.}
\label{asymf15}
\end{figure}

\begin{figure}
\vspace{0.5cm}
\caption[Intrinsic variables]{
Geometric intrinsic variables characterizing the shape of baryons.}
\label{variables}
\end{figure}

\clearpage
\begin{center}
Figure~\ref{harmosc}
\end{center}
\vspace{2cm}
\setlength{\unitlength}{0.9pt}
\begin{picture}(300,200)(-80,0)
\thinlines
\put (  0,  0) {\line(1,0){300}}
\put (  0,200) {\line(1,0){300}}
\put (  0,  0) {\line(0,1){200}}
\put (300,  0) {\line(0,1){200}}
\thicklines
\put ( 70, 60) {\line(1,0){20}}
\put ( 70,100) {\line(1,0){20}}
\put ( 70,140) {\line(1,0){20}}
\put (110,140) {\line(1,0){20}}
\put (150,140) {\line(1,0){20}}
\put (190,140) {\line(1,0){20}}
\put (230,140) {\line(1,0){20}}
\thinlines
\put ( 30, 60) {$n=0$}
\put ( 30,100) {$n=1$}
\put ( 30,140) {$n=2$}
\put ( 92, 60) {$0^+_{A_1}$}
\put ( 92,100) {$1^-_{E}$}
\put ( 92,140) {$2^+_{A_1}$}
\put (132,140) {$2^+_{E}$}
\put (172,140) {$1^+_{A_2}$}
\put (212,140) {$0^+_{A_1}$}
\put (252,140) {$0^+_{E}$}
\end{picture}

\clearpage
\begin{center}
Figure~\ref{vibrations}
\end{center}
\vspace{2cm}
\setlength{\unitlength}{1pt}
\begin{picture}(400,200)
\thinlines
\put ( 65, 50) {$u$--vibration}
\put ( 50,100) {\circle{5}}
\put (130,100) {\circle{5}}
\put ( 90,180) {\circle{5}}
\put ( 50,100) {\line ( 4, 3){40}}
\put (130,100) {\line (-4, 3){40}}
\put ( 90,180) {\line ( 0,-1){50}}
\thicklines
\put ( 50,100) {\vector(-4,-3){12}}
\put (130,100) {\vector( 4,-3){12}}
\put ( 90,180) {\vector( 0, 1){15}}
\thinlines
\put (195, 50) {$v$--vibration}
\put (180,100) {\circle{5}}
\put (260,100) {\circle{5}}
\put (220,180) {\circle{5}}
\put (180,100) {\line ( 4, 3){40}}
\put (260,100) {\line (-4, 3){40}}
\put (220,180) {\line ( 0,-1){50}}
\thicklines
\put (180,100) {\vector( 4,-3){12}}
\put (260,100) {\vector(-4,-3){12}}
\put (220,180) {\vector( 0, 1){15}}
\thinlines
\put (325, 50) {$w$--vibration}
\put (310,100) {\circle{5}}
\put (390,100) {\circle{5}}
\put (350,180) {\circle{5}}
\put (310,100) {\line ( 4, 3){40}}
\put (390,100) {\line (-4, 3){40}}
\put (350,180) {\line ( 0,-1){50}}
\thicklines
\put (310,100) {\vector(-1,-2){ 6}}
\put (390,100) {\vector(-1, 2){ 6}}
\put (350,180) {\vector( 1, 0){15}}
\end{picture}

\clearpage
\begin{center}
Figure~\ref{string}
\end{center}
\vspace{2cm}
\setlength{\unitlength}{0.9pt}
\begin{picture}(460,290)(0,0)
\thinlines
\put (  0,  0) {\line(1,0){460}}
\put (  0,290) {\line(1,0){460}}
\put (  0,  0) {\line(0,1){290}}
\put (460,  0) {\line(0,1){290}}
\thicklines
\put ( 30, 60) {\line(1,0){20}}
\put ( 30,100) {\line(1,0){20}}
\put ( 30,140) {\line(1,0){20}}
\put ( 70,100) {\line(1,0){20}}
\put ( 70,140) {\line(1,0){20}}
\put (110,140) {\line(1,0){20}}
\multiput ( 40,160)(0,5){5}{\circle*{0.1}}
\thinlines
\put ( 30, 15) {$(n_u,n_v+n_w)=(0,0)$}
\put ( 30, 35) {$K$=0}
\put ( 70, 35) {$K$=1}
\put (110, 35) {$K$=2}
\put ( 52, 60) {$0^+_{A_1}$}
\put ( 52,100) {$1^+_{A_2}$}
\put ( 92,100) {$1^-_{E}$}
\put ( 52,140) {$2^+_{A_1}$}
\put ( 92,140) {$2^-_{E}$}
\put (132,140) {$2^+_{E}$}
\thicklines
\put (160, 90) {\line(1,0){20}}
\put (160,130) {\line(1,0){20}}
\put (160,170) {\line(1,0){20}}
\put (200,130) {\line(1,0){20}}
\put (200,170) {\line(1,0){20}}
\put (240,170) {\line(1,0){20}}
\multiput (170,190)(0,5){5}{\circle*{0.1}}
\thinlines
\put (160, 15) {$(n_u,n_v+n_w)=(1,0)$}
\put (160, 35) {$K$=0}
\put (200, 35) {$K$=1}
\put (240, 35) {$K$=2}
\put (182, 90) {$0^+_{A_1}$}
\put (182,130) {$1^+_{A_2}$}
\put (222,130) {$1^-_{E}$}
\put (182,170) {$2^+_{A_1}$}
\put (222,170) {$2^-_{E}$}
\put (262,170) {$2^+_{E}$}
\thicklines
\put (290,150) {\line(1,0){20}}
\put (290,190) {\line(1,0){20}}
\put (290,230) {\line(1,0){20}}
\put (330,190) {\line(1,0){20}}
\put (330,230) {\line(1,0){20}}
\put (390,230) {\line(1,0){20}}
\multiput (300,250)(0,5){5}{\circle*{0.1}}
\thinlines
\put (300, 15) {$(n_u,n_v+n_w)=(0,1)$}
\put (290, 35) {$K$=0}
\put (330, 35) {$K$=1}
\put (390, 35) {$K$=2}
\put (312,150) {$0^+_E$}
\put (312,190) {$1^+_E$}
\put (352,190) {$1^-_{A_1 A_2 E}$}
\put (312,230) {$2^+_E$}
\put (352,230) {$2^-_{A_1 A_2 E}$}
\put (412,230) {$2^+_{A_1 A_2 E}$}
\end{picture}

\clearpage
\begin{center}
Figure~\ref{signature}
\end{center}
\vspace{2cm}
\large
\setlength{\unitlength}{1pt}
\begin{picture}(340,370)(0,0)
\thicklines
\put (130, 40) {\line(1,0){210}}
\put (130,350) {\line(1,0){210}}
\put (130, 40) {\line(0,1){310}}
\put (340, 40) {\line(0,1){310}}
\put (130, 80) {\line(1,0){5}}
\put (130,205) {\line(1,0){5}}
\put (130,330) {\line(1,0){5}}
\put (110, 80) {1.0}
\put (110,205) {1.5}
\put (110,330) {2.0}
\put ( 70,280) {M (GeV)}
\put (160, 40) {\line(0,1){5}}
\put (190, 40) {\line(0,1){5}}
\put (220, 40) {\line(0,1){5}}
\put (250, 40) {\line(0,1){5}}
\put (280, 40) {\line(0,1){5}}
\put (310, 40) {\line(0,1){5}}
\put (150, 20) {$5/2^-$}
\put (180, 20) {$3/2^-$}
\put (210, 20) {$1/2^-$}
\put (240, 20) {$1/2^+$}
\put (270, 20) {$3/2^+$}
\put (300, 20) {$5/2^+$}
\put (160,247) {\line(0,1){ 4}}
\put (160,249) {\circle{5}}
\put (190,208) {\line(0,1){ 4}}
\put (190,242) {\line(0,1){25}}
\put (190,210) {\circle{5}}
\put (190,255) {\circle{5}}
\put (220,210) {\line(0,1){ 9}}
\put (220,240) {\line(0,1){10}}
\put (220,214) {\circle{5}}
\put (220,242) {\circle{5}}
\put (250, 65) {\circle{5}}
\put (280,242) {\line(0,1){25}}
\put (280,260) {\circle{5}}
\put (310,249) {\line(0,1){ 4}}
\put (310,250) {\circle{5}}
\thicklines
\put (150,249) {\line(1,0){20}}
\put (180,221) {\line(1,0){20}}
\put (180,249) {\line(1,0){20}}
\put (210,221) {\line(1,0){20}}
\put (210,249) {\line(1,0){20}}
\put (240, 65) {\line(1,0){20}}
\put (270,262) {\line(1,0){20}}
\put (300,262) {\line(1,0){20}}
\multiput (150,237)(5,0){5}{\circle*{0.1}}
\multiput (180,204)(5,0){5}{\circle*{0.1}}
\multiput (180,236)(5,0){5}{\circle*{0.1}}
\multiput (210,195)(5,0){5}{\circle*{0.1}}
\multiput (210,214)(5,0){5}{\circle*{0.1}}
\multiput (240, 70)(5,0){5}{\circle*{0.1}}
\multiput (270,279)(5,0){5}{\circle*{0.1}}
\multiput (300,272)(5,0){5}{\circle*{0.1}}
\end{picture}
\normalsize

\end{document}